\documentclass[twocolumn]{aastex63}
\usepackage{amsmath}
\usepackage{hyperref}
\usepackage{graphicx}
\usepackage{booktabs}
\usepackage{diagbox}

\newcommand{\mbh}{\rm{$M_{\rm{BH}}$}}
\newcommand{\ha}{\rm{H$\alpha$}}
\newcommand{\hb}{\rm{H$\beta$}}

\usepackage{xcolor}
\usepackage{bold-extra}

\submitjournal{ApJ}

\shorttitle{Pilot Study for the CSST Slitless Spectroscopic Quasar Survey}
\shortauthors{Pang et al.}

\begin{document}

\title{A Pilot Study for the CSST Slitless Spectroscopic Quasar Survey Based on Mock Data}

\author[0009-0005-3823-9302]{Yuxuan Pang}
\affiliation{Department of Astronomy, School of Physics, Peking University, Beijing 100871, People's Republic of China}
\affiliation{Kavli Institute for Astronomy and Astrophysics, Peking University, Beijing 100871, People's Republic of China}

\author[0000-0002-7350-6913]{Xue-Bing Wu}
\affiliation{Department of Astronomy, School of Physics, Peking University, Beijing 100871, People's Republic of China}
\affiliation{Kavli Institute for Astronomy and Astrophysics, Peking University, Beijing 100871, People's Republic of China}


\author[0000-0002-0759-0504]{Yuming Fu}
\affiliation{Leiden Observatory, Leiden University, P.O. Box 9513, NL-2300 RA Leiden, The Netherlands}
\affiliation{Kapteyn Astronomical Institute, University of Groningen, P.O. Box 800, NL-9700 AV Groningen, The Netherlands}

\author[0000-0002-0792-2353]{Rui Zhu}
\affiliation{Department of Astronomy, School of Physics, Peking University, Beijing 100871, People's Republic of China}
\affiliation{Kavli Institute for Astronomy and Astrophysics, Peking University, Beijing 100871, People's Republic of China}

\author[0000-0001-8879-368X]{Bing Lyu}
\affiliation{Department of Astronomy, School of Physics, Peking University, Beijing 100871, People's Republic of China}
\affiliation{Kavli Institute for Astronomy and Astrophysics, Peking University, Beijing 100871, People's Republic of China}

\author[0000-0001-8803-0738]{Huimei Wang}
\affiliation{Department of Astronomy, School of Physics, Peking University, Beijing 100871, People's Republic of China}
\affiliation{Kavli Institute for Astronomy and Astrophysics, Peking University, Beijing 100871, People's Republic of China}

\author[0000-0003-0174-5920]{Xiaotong Feng}
\affiliation{Department of Astronomy, School of Physics, Peking University, Beijing 100871, People's Republic of China}
\affiliation{Kavli Institute for Astronomy and Astrophysics, Peking University, Beijing 100871, People's Republic of China}

\begin{abstract}

The wide survey of the Chinese Space Station Telescope (CSST) will observe a large field of 17,500 $\text{deg}^2$. The GU, GV, and GI grism observations of CSST will cover a wavelength range from 2550 to 10000$\rm{\AA}$ at a resolution of $R\sim 200$ and a depth of about 22 AB magnitude for the continuum. In this paper, we present a pipeline to identify quasars and measure their physical properties with the CSST mock data. We simulate the raw images and extract the one-dimensional grism spectra for quasars, galaxies, and stars with the r-band magnitudes of $18<\text{m}_{\text{r}}<22$ using the CSST Cycle 6 simulation code\footnote{\url{https://csst-tb.bao.ac.cn/code/csst_sim/csst-simulation}\label{footnote1}}. Using a convolution neural network, we separate quasars from stars and galaxies. We measure the redshifts by identifying the strong emission lines of quasars. We also fit the 1D slitless spectra with QSOFITMORE to estimate the black hole masses and Eddington ratios. Our results show that the CSST slitless spectroscopy can effectively separate quasars with redshifts $z=0-5$ from other types of objects with an accuracy of 99\%. Among those successfully classified quasars, 90\% of them could have precise redshift measurements with $\sigma_{\mathrm{NMAD}}=0.002$. The scatters of black hole masses and Eddington ratios from the spectral fittings are 0.13 and 0.15 dex, respectively. The metallicity diagnosis line ratios have a scatter of 0.1-0.2 dex. Our results show that the CSST slitless spectroscopy survey has the potential to discover about 0.9 million new quasars and provide important contributions to AGN science and cosmology.

\end{abstract}
\keywords{Active galactic nuclei (16) --- Galactic and extragalactic
astronomy (563) --- Quasars (1319) --- Spectroscopy(1558) --- Supermassive black holes (1663)}

\section{Introduction} \label{sec:intro}
As luminous Active Galactic Nuclei (AGNs), quasars (QSOs) can emit immense energy from an intense accretion process, making them pivotal objects in modern astronomy. The existence of Supermassive Black Holes (SMBHs) at the center of QSOs in the early universe challenges our SMBH growing process \citep{2011Natur.474..616M,2015Natur.518..512W,2020ApJ...897L..14Y}. The potential explanation includes higher accretion rate or larger black hole seed mass \citep[e.g.,][]{2020ARA&A..58...27I,2021NatRP...3..732V}. Observations spanning multiple wavelengths, including Spectral Energy Distributions (SED) and long-term variability, have developed a comprehensive, unified model explaining the dynamics and structure of QSOs' accretion disks, broad line regions, and dusty torus \citep{1993ARA&A..31..473A, 1995PASP..107..803U,2015ARA&A..53..365N}. Furthermore, statistical analyses of QSOs and their host galaxies have confirmed the well-known association between supermassive black holes and their galactic hosts \citep[e.g.,][]{2005Natur.433..604D, 2013ARA&A..51..511K, 2020ApJ...899..112S}.

Shining through the significant distances, QSOs are ideal back-light to study the Interstellar Medium (ISM), Circumgalactic Medium (CGM), and Intergalactic Medium (IGM) within the foreground galaxies and their surrounding clusters \citep[e.g.,][]{1986MNRAS.218P..25R, 2006ApJS..165....1T, 2015ApJ...809..163A, 2016ApJS..226...25L,2022NatAs...6..850J}. An evenly distributed, completed QSO catalog could serve as a celestial reference frame for sky surveys \citep[e.g.,][]{2018A&A...616A..14G, 2022A&A...667A.148G}. QSO catalogs are also very helpful in mapping the large-scale structure of the universe \citep[e.g.,][]{2013AJ....145...10D, 2017AJ....154...28B, 2023arXiv230915814C}. Recently, some studies are also trying to do the cosmological parameter constraints through a more completed Gaia all sky QSO catalog \citep[][]{2023JCAP...11..043A}. Using high-resolution spectra of the bright QSOs, the expansion history of the universe could be determined via the Sandage test \citep[][]{2023MNRAS.522.2019C}.

Previous QSO samples are dominated by the Sloan Digital Sky Survey (SDSS) observations; the sixteen data release contains 750,414 QSOs that have been spectroscopically identified from SDSS-I to SDSS-IV \citep[][]{2020ApJS..250....8L}. Recently, the Dark Energy Spectroscopic Instrument (DESI) early data release also included about 90k new QSOs \citep[][]{2023arXiv230606308D, 2024arXiv240403621J}. MILLIQUAS v8 catalog combines QSOs from the literature before July 2023, including 907k type-I QSOs \& AGNs and 66,026 high-confidence ($\sim$99\% likelihood) radio/X-ray associated quasar candidates \citep{2023OJAp....6E..49F}. Gaia satellite observations also build up an all-sky catalog containing several million candidates of QSOs \citep[][]{2024ApJ...964...69S, 2024ApJS..271...54F}. However, we still need an all-sky, high-precision and complete QSO sample to study cosmology at higher redshift, get the best astrometric reference frame, and evaluate the completeness of different QSO selection methods.

The Chinese Space Station Telescope (CSST) is a two-meter space telescope planned to launch in 2026. With a large Field of View (FOV) of 1.1 $\text{deg}^2$, high spatial resolution of 0.15 arcsec, and broad wavelength coverage from 2550$\rm{\AA}$ to 10000$\rm{\AA}$, the wide field survey will provide seven band imaging (hereafter NUV, u, g, r, i, z, and y band) and slitless spectroscopic data from GU (2550$\rm{\AA}$ to 4000$\rm{\AA}$), GV (4000$\rm{\AA}$ to 6000$\rm{\AA}$) and GI (6000$\rm{\AA}$ to 10000$\rm{\AA}$) grisms over 17,500 $\text{deg}^2$ in ten years \citep[][]{2011SSPMA..41.1441Z,2018cosp...42E3821Z,2021ChSBu...66...1290}. CSST observations could help us obtain a large, complete QSO sample, spanning a redshift range of 0--6 through photometry and slitless spectroscopy. Such a complete sample covers a larger sky area than SDSS and DESI while having better spectral resolution and image quality than Gaia. In addition, CSST's deep multi-band imaging allows us to study the host galaxy properties and the coevolution with SMBHs at higher redshift compared to the ground-based surveys.

In this paper, we focus on studying the performance of QSO survey from the CSST slitless spectroscopy, specifically in QSO selection accuracy, redshift measurement precision, and the estimation of physical parameters. The redshift range of QSOs in this work are $z=0-5$, where C~\textsc{IV} line is still observed in the GI grism, which help us to distinguish QSOs from the Ly$\alpha$ emitters. The paper is organized as follows. In Section \ref{sec: data}, we detail the CSST Cycle 6 simulation procedure, including the generation of mock catalogs for QSOs, galaxies and stars, as well as the extraction of 1D slitless spectra. In section \ref{sec: distinguish}, we introduce a neural network to distinguish QSOs from galaxies and stars. In Section \ref{sec: redshift}, we determine the precise QSO redshifts through emission line identifications. In Section \ref{sec: fitting}, we present the spectral fitting procedure to compute various spectral properties. Using the fitting results, we estimate the black hole mass, the Eddington ratio, and the metallicity diagnostics. Finally, we discuss the accuracy of wavelength and flux calibrations and the real observation strategy influence in Section \ref{sec: discuss}. In this work, we adopt a flat $\Lambda$CDM cosmology with parameters $\Omega_m$ = 0.30, $\Omega_{\Lambda}$  = 0.7 and $h_0$ = 70 km $\text{s}^{-1}$ $\text{Mpc}^{-1}$.

\section{Simulation Data} \label{sec: data}
In this section, we introduce the CSST simulation code used in this work and the input mock data catalog, Spectral Energy Distributions (SED), and other physical parameters. 

\subsection{CSST Cycle 6 Simulation Procedure}
The CSST Cycle 6 simulation procedure is a precise observation simulation framework (see more details in the CSST simulation platform\footref{footnote1}). In addition to standard simulation factors such as the read-out noise, the simulation accounts for other realistic observational effects including optical and structural deviations from the ideal design due to the telescope engineering structures and point spread function (PSF) distributions in different bands derived from the laboratory settings.

\begin{figure*}[htbp]
    \centering
    \includegraphics[width=1\textwidth]{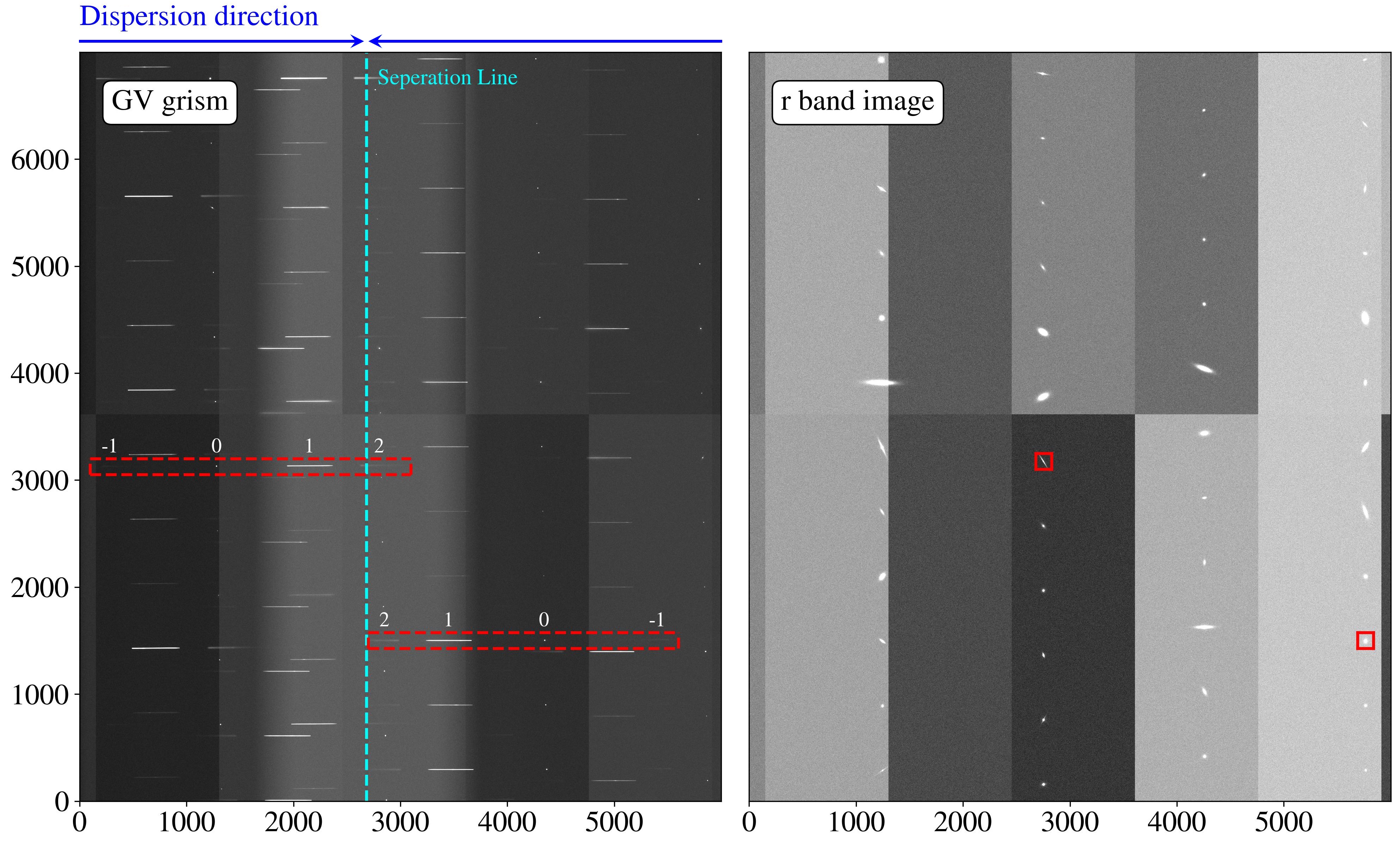}
    \caption{Illustration of the CSST Cycle 6 simulation results. \textbf{Left panel:} Results from the GV grism simulation. Due to the design of the dispersing instrument, the dispersion direction differs on the opposite sides of CCD (as indicated by the blue arrows at the top of panel, with the cyan dashed line marking the dividing line). The background distribution is uneven across the field.  The positions of the 0, $\pm$1, and 2-order spectra for two sources with different dispersion directions are labeled. \textbf{Right panel:} Image simulation of the r-band in the same sky area. The square-shaped background pattern is caused by the different biases and gains of the 16 channels. The direct images of two sources highlighted in the left panel are marked in red square.
    }
    \label{fig: simu_example}
\end{figure*}

The inputs of the simulation code include the satellite's pointing location, the specific chip number to determine the photometry image or grism spectroscopy, and an input catalog containing the spatial coordinates and SED. Additional parameters such as the Sérsic index and half-light radius are considered for galaxies. Figure \ref{fig: simu_example} illustrates an example of the simulated results for the GV grism and r band imaging.

During the simulation, we consider various factors, including the different bias and gain settings for each of the 16 channels, the flat fielding effect, the Photo Response Non-Uniformity (PRNU) patterns, the non-linear responses, and the charge transfer efficiency (CTE) trails. We exclude the influences of cosmic rays and dead pixels since these effects remove specific pixels directly from the CCD, reducing the final signal-to-noise ratio (SNR) but not significantly impacting the spectrum extraction procedure.

\subsection{Basic Settings}
The simulation centers around the coordinates RA, DEC = 150.0\textdegree, 2.0\textdegree, at a high galactic latitude region. To reduce contamination from nearby sources, the simulated sources are deliberately distributed sparsely within the CCD region\textemdash approximately 30--45 times less dense than real star fields, depending on galactic latitude. The exposure time for GU, GV, and GI grism is 4*150s, the same as the CSST wide-field survey. We exclude all the sources with the r-band apparent magnitude $\text{m}_{\text{r}}>22$ to align with the effective sensitivity for CSST's GU, GV, and GI grisms.

The input wavelength for all mock data is standardized from 1500$\rm{\AA}$ to 12500$\rm{\AA}$ at a spectral resolution of 4000. For stars and QSOs, the reference image is the exact PSF pattern at the CCD location. A single sérsic model is convolved with the PSF pattern to create the reference image for galaxies.

\subsubsection*{Galaxy Input Catalog}
\begin{figure}[htbp]
    \centering
    \includegraphics[width=0.47\textwidth]{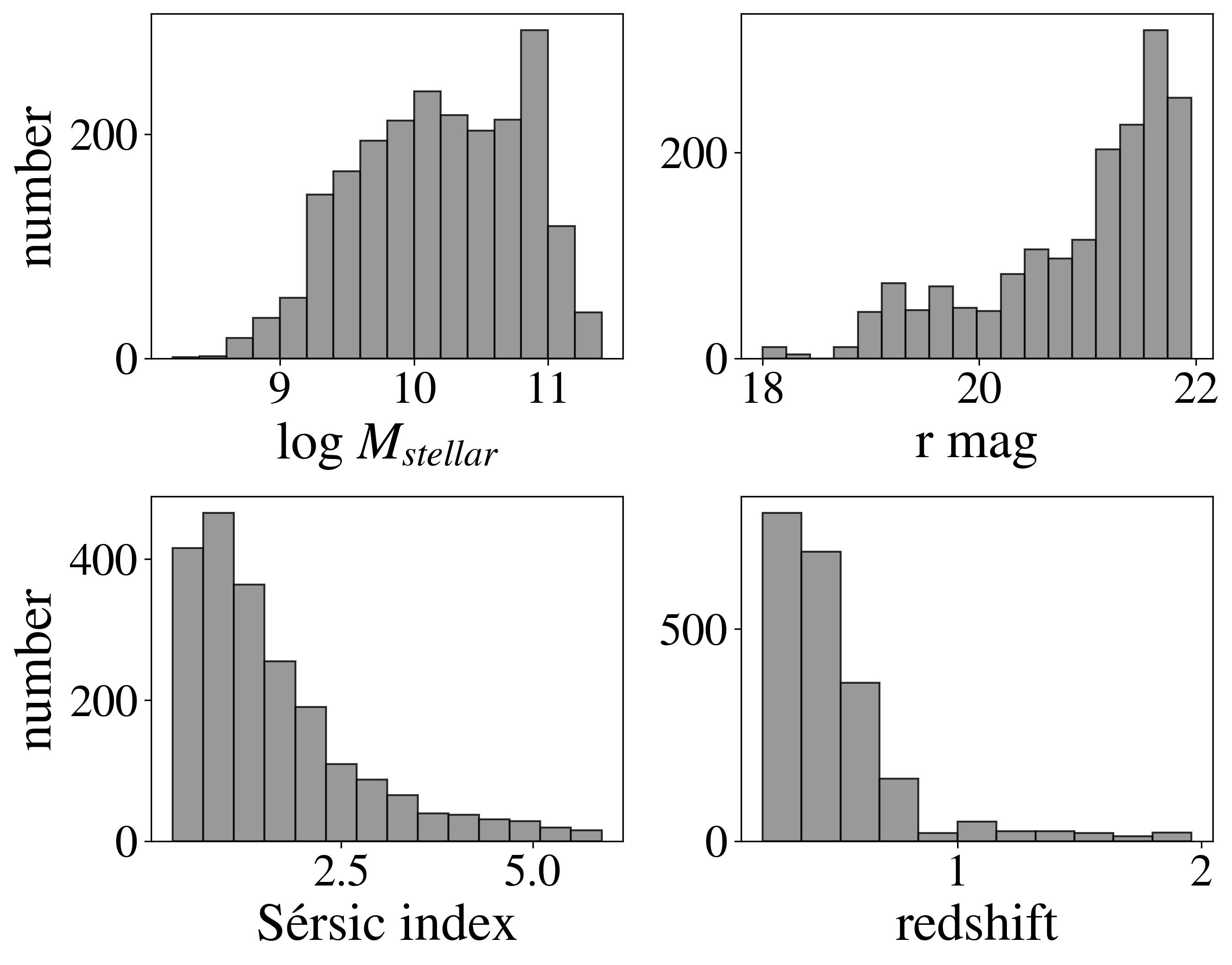}
    \caption{The typical parameter distributions of the input galaxies, from top-left to bottom-right: stellar mass ($\text{M}_{\text{stellar}}$), r-band apparent magnitude ($\text{m}_{\text{r}}$), Sérsic index, and redshift.}
    \label{fig: gal_para}
\end{figure}

We gather all JWST mock data\footnote{\url{https://fenrir.as.arizona.edu/jwstmock/}} \citep[][]{2018ApJS..236...33W}, including the SEDs and morphology information for galaxies spanning an area of 1210 $\text{arcmin}^2$. The redshift distribution of these galaxies is depicted in Figure \ref{fig: gal_para}, predominantly falling within the redshift range of 0-0.9. Figures \ref{fig: gal_para} also illustrates the distribution of effective radius, Sérsic index, and ellipticity within the galaxy input catalog.

\subsubsection*{QSO Input Catalog}
The modeling spectra for QSOs are simulated from the SimQSO package\footnote{\url{https://github.com/imcgreer/simqso?tab=readme-ov-file}} \citep{2021ascl.soft06008M}, randomly selected from the simulated spectra database spanning a redshift range of 0--5. The absolute magnitudes and emission line strength distributions in our SimQSO simulation are the same as the SDSS eBOSS luminosity function and statistic analysis \citep[][]{2018A&A...613A..51P}. For QSOs with redshifts below 2, we associate each QSO with a host galaxy. To account for the potential evolution of the relationship between QSOs and their host galaxies with redshift, the stellar mass ($\text{M}_{\text{stellar}}$) of QSO host galaxy is calculated by $\text{M}_{\text{BH}}$ using the $\text{M}_{\text{BH}}$\textendash$\text{M}_{\text{stellar}}$ relation provided by the TRINITY model \citep{2024MNRAS.529.2777Z}.
This model integrates observational results across a wide redshift range to describe the evolution of the $\text{M}_{\text{BH}}$\textendash $\text{M}_{\text{stellar}}$ relationship and aligns with the M\textendash $\sigma$ relation in the local universe. Other properties of the host galaxy are then randomly selected in the JWST mock data after the stellar mass constraint. During the simulation, the host galaxy is positioned at the exact location of QSO to simulate its influence on the slitless spectra.

\begin{figure}[htbp]
    \centering
    \includegraphics[width=0.47\textwidth]{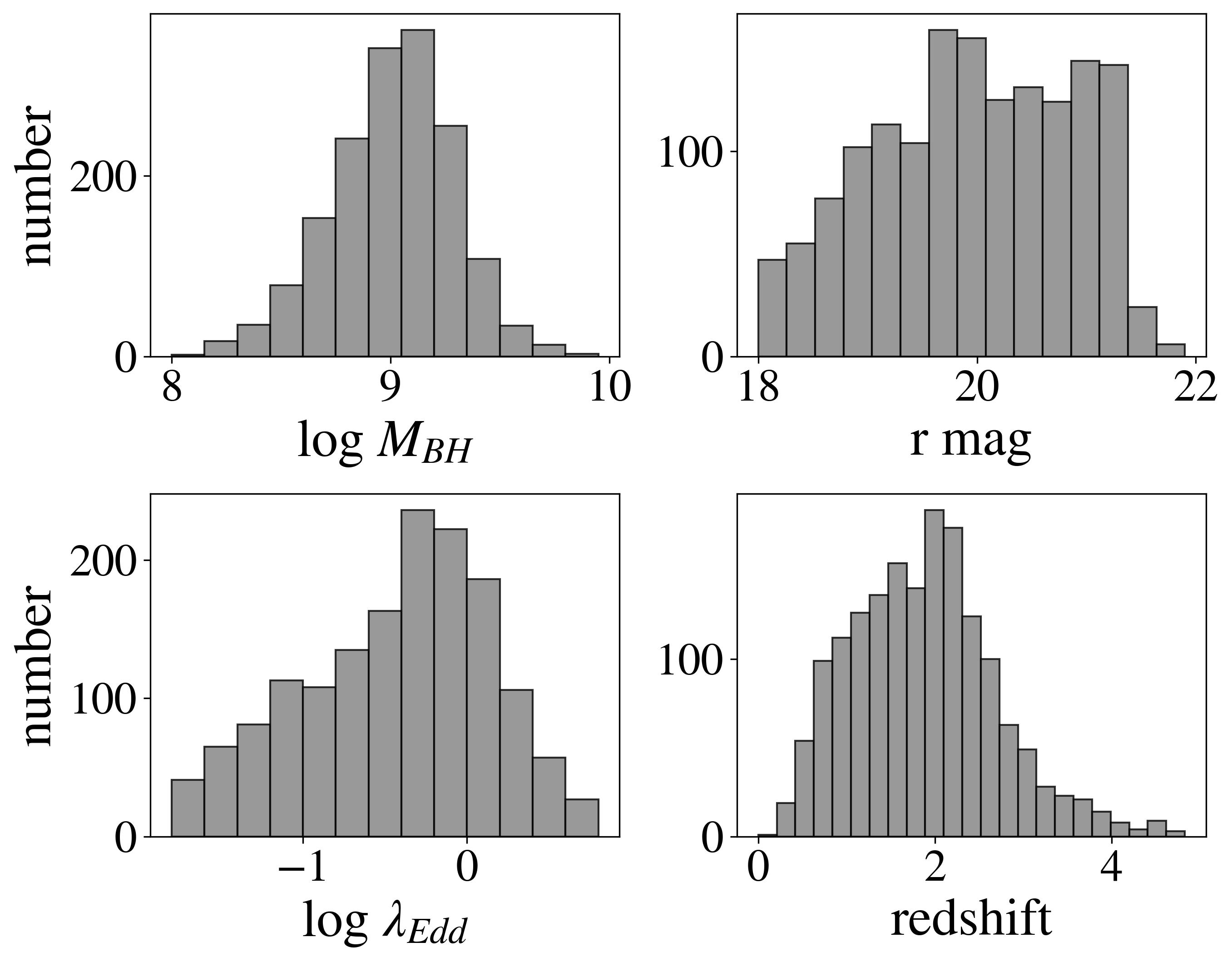}
    \caption{The typical parameter distributions of the input QSOs, from top-left to bottom-right: supermassive black hole mass ($\text{m}_{\text{BH}}$), r-band apparent magnitude ($\text{m}_{\text{r}}$), Eddington ratio ($\lambda_{\rm{Edd}}$), and redshift.}
    \label{fig: agn_para}
\end{figure}

\subsubsection*{Star Input Catalog}
The SEDs for stars are derived from two parts. The first part, covering stars from type B to early-type M, is based on the stellar spectra provided by \citet{2018A&A...618A..25A}. Additionally, we incorporate the SDSS spectra from the MaNGA Stellar Library \citep{2022ApJS..259...35A} for the late-type M dwarfs. We then convolve the stellar SED to get their g band mangitude and follow the same number density function in g band as in Figure 4 of \citet{2023SCPMA..6619511C}.

Our final input catalog comprises 1399 star-forming galaxies, 763 quiescent galaxies, 1646 QSOs, and 1468 stars. The catalog is mock-observed by 765 CCD images in 85 satellite orbits.

\subsection{Data Reduction}  \label{subsec: reduction}
To reduce the instrument effects, we simulate the bias, dark current, and flat field data for each observation and utilize the ccdproc package\footnote{\url{https://ccdproc.readthedocs.io/en/latest/}} to rectify these effects. From the celestial coordinates (RA, DEC) of the sources, we compute their ideal positions on each grism CCD. We then apply the distortion function specific to each CCD to determine the observed location of the source accurately. 

Following the steps above, using the configuration data akin to aXe software \citep{2009PASP..121...59K}, we cut out the zero and first-order slitless spectra in each image. To ensure most of the flux is included even for the extended galaxies, a cut aperture of 100 pixels (about 7.4 arcsec) is imposed in this step. Additionally, the wavelength estimation is performed for the pixel array in the dispersion direction. Subsequently, the background estimation is employed to mitigate the local background effects by the Python SEP package\footnote{\url{https://sep.readthedocs.io/en/v1.1.x/}}. The one-dimensional spectra are extracted from the resultant first-order cut-out image with a flexible aperture determined by the zero-order image. Finally, the sensitivity curve is applied to the one-dimensional spectra. 

\begin{figure*}[htbp]
    \centering
    \includegraphics[width=0.7\textwidth]{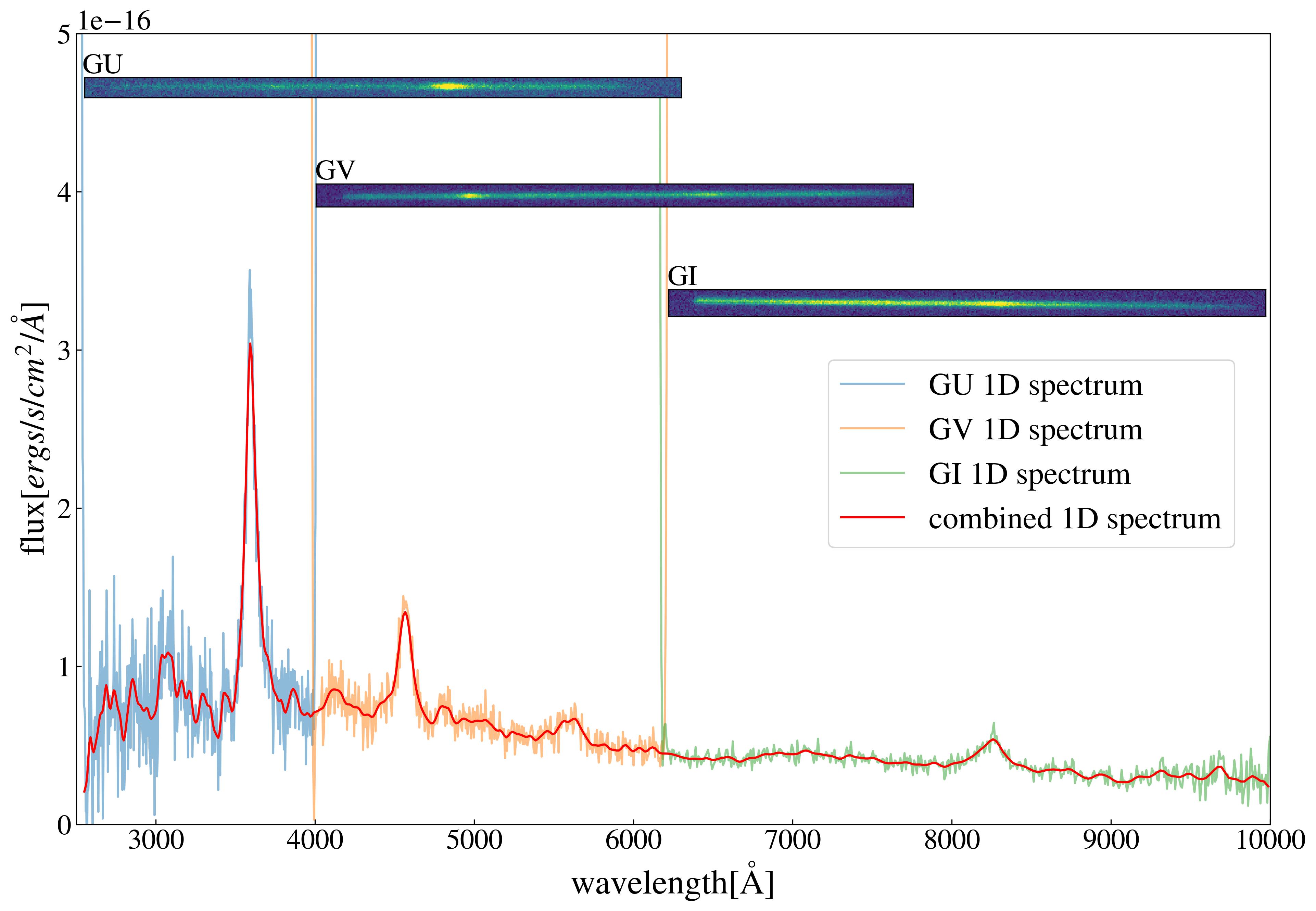}
    \caption{The extracted 2D and 1D spectrum of a simulated QSO. The panels on the top show the 2D grism images of GU, GV, and GI, respectively. The blue, orange, and green line shows the extracted 1D spectrum of each grism. The red line shows the final combined 1D spectrum.}
    \label{fig: 1d_spec}
\end{figure*}

Since the CSST imaging exposure time is the same as slitless spectroscopy, the imaging data is approximately two magnitudes deeper than the grism observations. In principle, this allows the 1D spectra extraction for all simulated sources. However, spectra with low SNR lack enough information and may introduce noise that contaminates the training of classification models. On the other hand, setting a high SNR threshold could lead to the exclusion of some quasars, reducing the completeness of the final sample. Balancing these considerations, we abandon the extracted spectra in GU, GV, and GI bands with SNR \textless~5, and also reject sources with total SNR \textless~9 in the combined three band spectra. With this criterion, we extracted the spectra for 1,645 QSOs, 1,255 galaxies, and 1,468 stars from all simulated sources. A portion of galaxies with diffuse morphologies have relatively low SNR in their spectra, resulting in their exclusion. Figure \ref{fig: 1d_spec} shows the 2D and 1D spectrum extraction results for a typical QSO.

\section{QSO Classification from 1D Slitless Spectra} \label{sec: distinguish}
We employ a neural network (NN) to train a machine learning classification model, which allows us to reclassify the extracted spectra into QSOs, stars, and galaxies \citep[][]{2018arXiv180809955B}. To focus on distinguishing the cases of the Balmer Break of low-redshift galaxies and high-redshift QSOs, we further subdivide the QSO sample into low-z and high-z classes by redshift 2, where Lyman $\alpha$ emission lines entirely move into the GU grism.

The dataset used in this section consists of the spectra of 1645 QSOs, 1255 galaxies, and 1468 stars, extracted as outlined in Section \ref{subsec: reduction}. We randomly partition the spectral sample into the training and validation sets using a 4:1 ratio. Initially, for both the training and the validation set, we combine the extracted GU, GV, and GI spectra. We apply the sigma-clipping method (\verb|astropy.sigmaclipping|) at the overlap regions among the GU, GV, and GI grisms to remove any 3-sigma outlier data points in the grism low-resolution region. Subsequently, we resample the combined spectra to the range [2550$\rm{\AA}$, 10000$\rm{\AA}$] with a fixed spectral resolution of 400 (\verb|specutils.resample|) and calibrate the spectra to absolute fluxes.

\begin{figure}[htbp]
    \centering
    \includegraphics[width=0.47\textwidth]{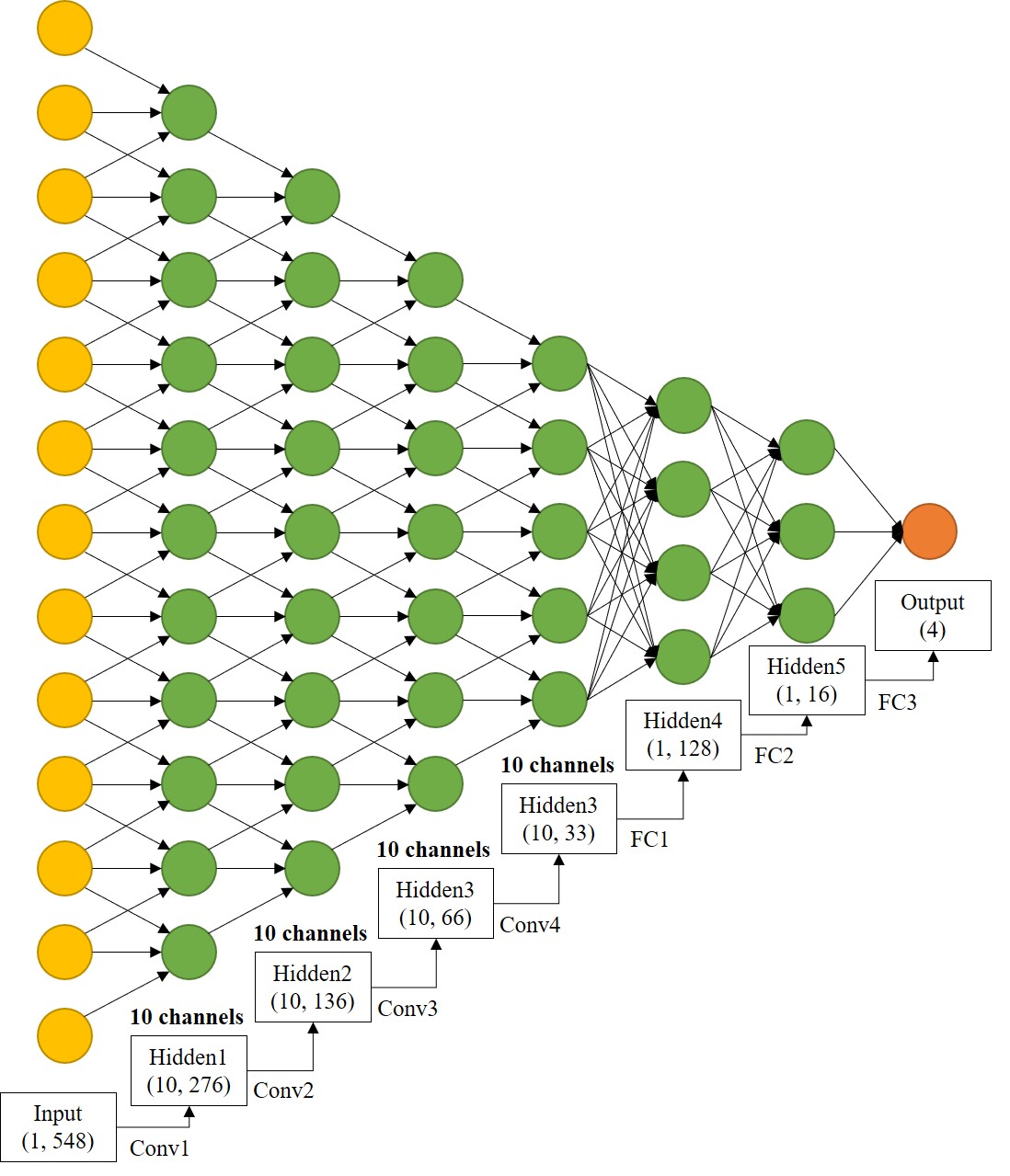}
    \caption{Schematic diagram of the RegNet architecture, designed to classify the 1D extracted spectra. This diagram shows the process of a single spectrum with 548 points passing through the network and yielding the classification probability. Only a small fraction of the input and hidden neurons are plotted for simplicity.}
    \label{fig: nn_structure}
\end{figure}

The RegNet \citep{2020arXiv200313678R} architecture consists of four convolutional layers followed by three fully connected linear layers, culminating in a 4 class output consisting of low-z QSOs, high-z QSOs, galaxies, and stars. Each input spectrum contains 548 data points (neurons) and is scaled to a mean value of 1 with a standard error of 0.5; the missing value of the spectra is filled with number -1. Each convolutional layer has ten channels and a kernel size of 5, and the output of which goes through a ReLU activation function and a MaxPool function with a kernel size of 2. Each fully connected layer connects all neurons from the last layer to 128, 100, and 16 neurons and applies a ReLU activation function to the output; the final output layers contain four neurons, representing each class's probability, respectively. Figure \ref{fig: nn_structure} shows a schematic diagram of the neural network structure.

The NN model is trained in the shuffled batches, each containing ten spectra. With the default parameters of the Adam optimizer (\verb|torch.optim.Adam|), we train the RegNet model for 200 epochs. The multi-class log loss $L_{\log}(Y, P)$ (also known as cross-entropy loss) is used as the objective function to be minimized during the model training and hyperparameter optimization. For a classification task with $K$ classes and $N$ samples, let the true label of a sample $i$ be encoded as a binary indicator $y_{i,k} \in \{0,1\}$, and $y_{i,k} = 1$ when a sample $i$ has label $k$. A probability estimate is defined as $p_{i,k} = \text{Pr}(y_{i,k} = 1)$. Let $P$ be the matrix of probability estimates and $Y$ be the encoded label matrix. The log loss of the whole set is the negative log-likelihood of the classifier given the true labels:

\begin{equation}
    L_{\log}(Y, P) = - \frac{1}{N} \sum_{i=0}^{N-1} \sum_{k=0}^{K-1} y_{i,k} \ln p_{i,k}.
\end{equation}

Other metrics used to evaluate the model performance are: balanced accuracy, precision, recall, F1, and Matthews correlation coefficient (MCC). For binary classification problems, with the true positive denoted as TP, true negative as TN, false positive as FP, and false negative as FN, the five metrics are defined as:

\begin{equation}
    \mathrm{balanced\ accuracy = \frac{1}{2}\left( \frac{TP}{TP + FN} + \frac{TN}{TN + FP}\right )}
\end{equation}
\begin{equation}
    \mathrm{precision = \frac{TP}{TP+FP}}
\end{equation}
\begin{equation}
    \mathrm{recall = \frac{TP}{TP+FN}}
\end{equation}
\begin{equation}
    F_{1} = \mathrm{2\times \frac { precision \times recall}{precision + recall}}
\end{equation}
\begin{equation}
\begin{split}
    & \rm MCC \\
    & =\mathrm{ \frac{TP \times TN - FP \times FN}{\sqrt{(TP+FP)(TP+FN)(TN+FP)(TN+FN)}}}.
\end{split}
\end{equation}

In the case of a multi-class problem, the classification task is treated as a collection of binary classification problems, one for each class. The five metrics above can be calculated for each binary classification problem (each class). The metrics of the multi-class problem are the average metrics of all classes. We adopt the functions \verb|balanced_accuracy_score|, \verb|precision_score|, \verb|recall_score|, \verb|f1_score|, and \verb|matthews_corrcoef| of the \verb|sklearn.metrics| in the scikit-learn module to calculate the metrics for the four-class classification problem in this work. When calculating the precision, recall, and $F_{1}$, the `\verb|weighted|' strategy is used, in which the score of each class is weighted by its fraction in the true data sample.

\begin{figure}[htbp]
    \centering
    \includegraphics[width=0.40\textwidth]{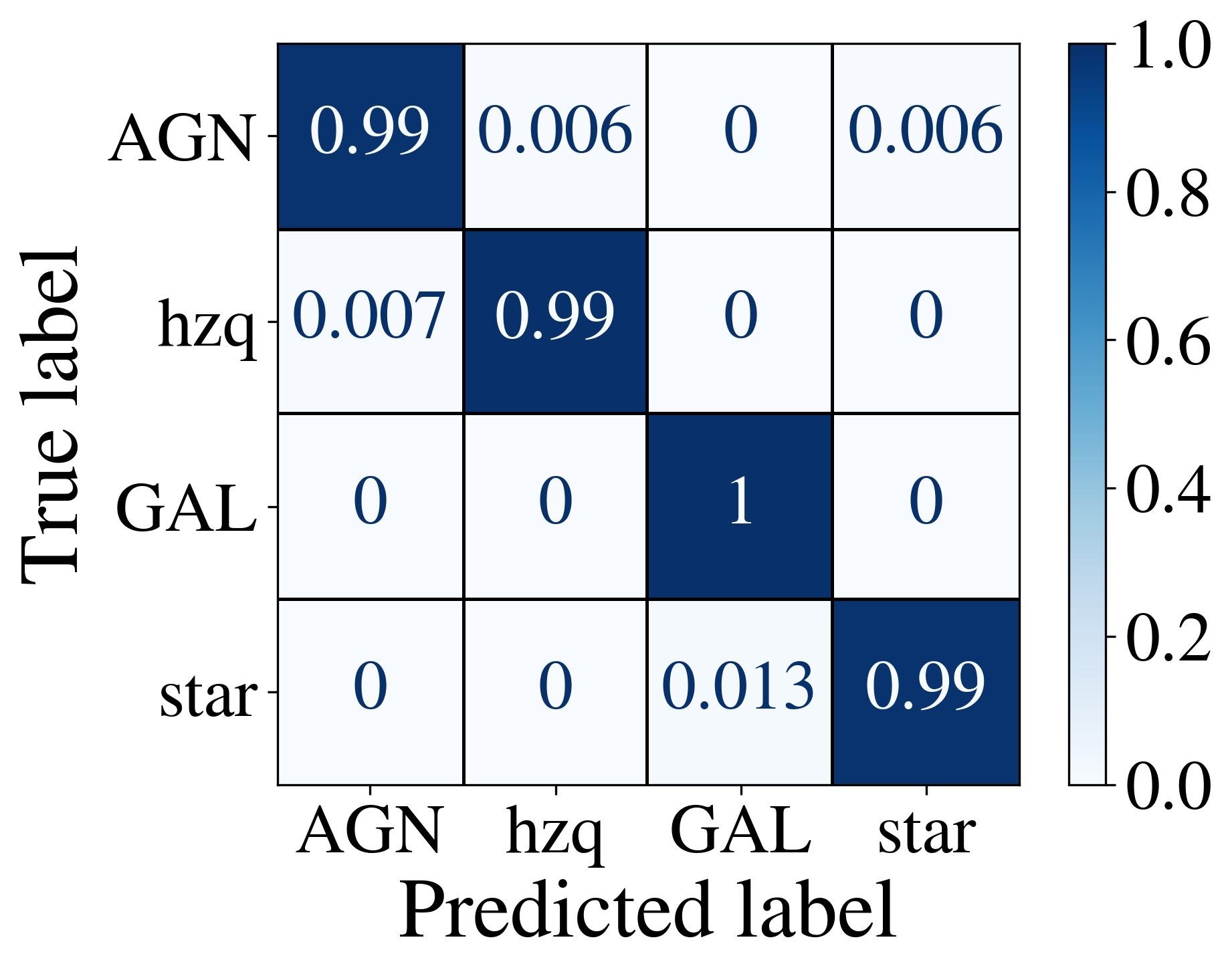}
    \caption{The confusion matrix of the extracted 1D spectra NN classification results in four classes. The label `AGN', `hzq', `GAL' and `star' class represent the low-z QSO, high-z QSO, galaxy and star, respectively.}
    \label{fig: confusion-matrix}
\end{figure}

The optimal model is from the epoch with the lowest validation loss. The voting results of 15 NN give the final class. On the validation set, the balanced accuracy, precision, recall, F1, and Matthews correlation coefficient (MCC) of the NN model are shown in the first line in Table \ref{tab: accuracy_class}. Figure \ref{fig: confusion-matrix} shows the spectra's confusion matrix of the four-class classification. 

Although the majority of sources are accurately classified, a few cases still have confusion. We examined the failure cases of our model. For low-z QSOs, a strong C~\textsc{IV} line can lead to the misclassification as Ly $\alpha$, causing confusion with high-z QSOs. Low SNR of the continuum for a few high-z QSOs may also be misclassified as low-z QSOs. Late-type stars and quiescent galaxies may show a similar continuum in the GV and GI bands and in the case of low SNR in GU bands, leading to misidentification. These cases suggest that our current model is not strong enough in the detection of emission and absorption line features. In the future, by expanding the training sample, we can build a deeper neural network to distinguish these features more accurately and improve the success rate.

We use the quasar luminosity functions from \citet{2018A&A...613A..51P} and \citet{2020MNRAS.495.3252S} to estimate that approximately 1.7 million QSOs in the redshift range of 0 to 5 reside within the survey area of the CSST wide-field slitless spectroscopy. The QSO selection completeness for CSST grism observations depends on three factors: spectral extraction completeness, contamination rate, and classification completeness.

Since CSST imaging data is two magnitudes deeper than its slitless spectroscopy observations, all sources down to 22 AB magnitude can reliably be identified in the muti-band image, and the SNR cut in Section \ref{subsec: reduction} only removes one out of 1646 simulated QSOs, ensuring the high spectral extraction completeness. According to \citet{10.1093/mnras/stae157}, about 10\% of sources are expected to experience severe contamination (more than 40\% of total flux), while another 20\% may face moderate contamination (10\%-40\% of total flux) in galaxy cluster fields observed with the CSST grism. Using the 3D-HST G800L observations of sources brighter than 22 AB magnitude in F814W photometry \citep{2016ApJS..225...27M}, we estimate that the contamination rate in CSST deep fields is likely to be below 5\%. Therefore, we expect the overall contamination rate for QSO selection with the CSST grism to be within 10\%. As shown in Table \ref{tab: accuracy_class}, the recall of our neural network (NN) model for QSO classification exceeds 99\%, indicating very high completeness in spectral classification.

Taking all these factors into account, we estimate the overall completeness of QSO selection for CSST to be approximately 90\%. Based on this completeness, over 1.5 million QSOs will be detected and classified by the CSST wide-field slitless spectroscopy survey. After excluding QSOs already identified by SDSS \citep[DR16Q,][]{2020ApJS..250....8L}, GAIA (CatNorth, \citealp{2024ApJS..271...54F}; CatSouth, Fu et al. in prep), and DESI \citep[EDR,][]{2023arXiv230606308D} in the overlap regions, about 0.9 million new QSOs are expected to be identified.

\section{Redshift Measurement}\label{sec: redshift}
Subsequent to the QSO classification, we build up an automatically measurement of the redshift for the spectra class of `QSO'. Since the broad wavelength span combines the GU, GV, and GI spectra, detecting and identifying the robust emission line features is natural to get precise redshift results. Thus, our procedure plans to find the emission line peaks and identify them. 

\begin{figure}[htbp]
    \centering
    \includegraphics[width=0.47\textwidth]{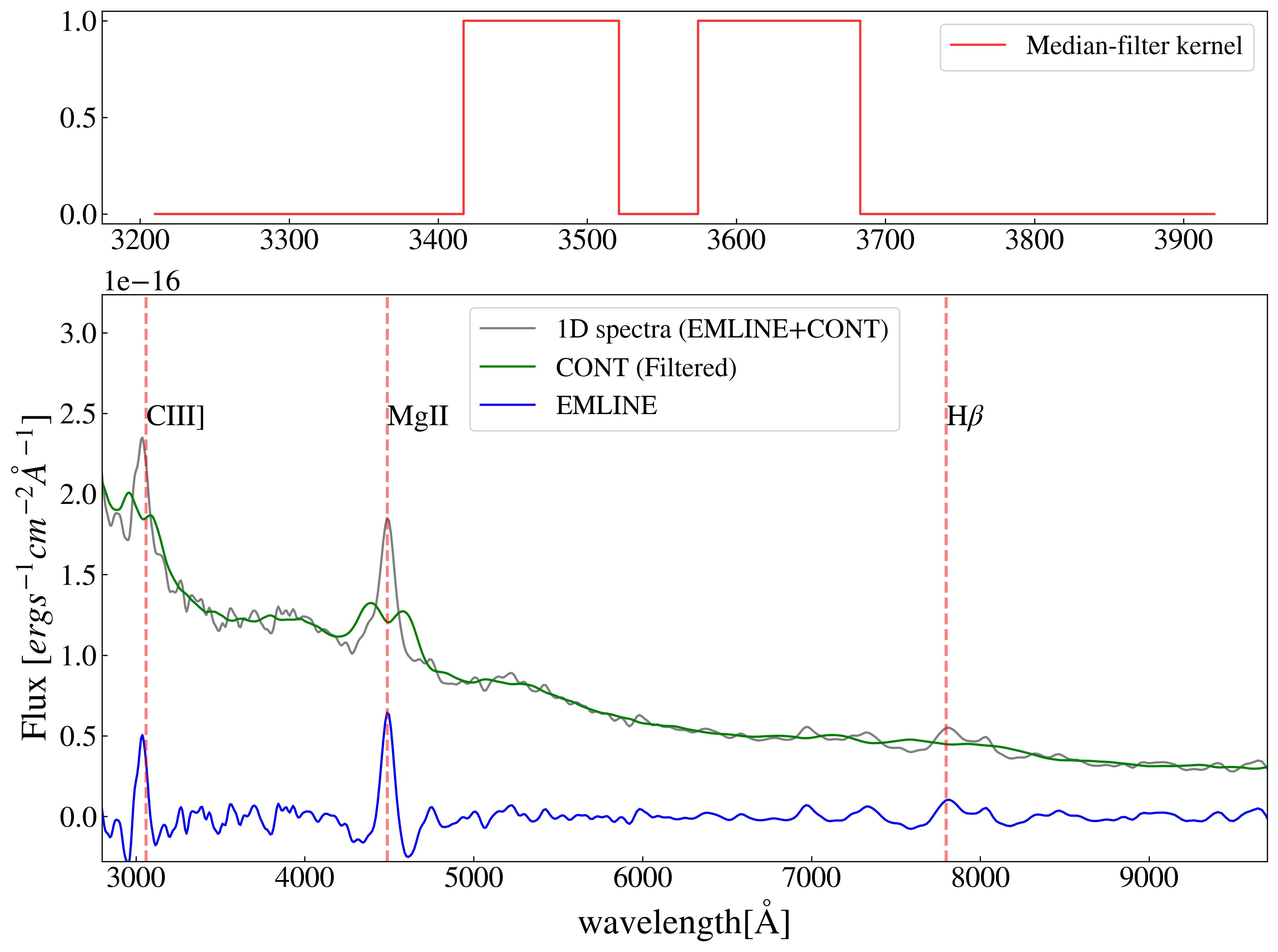}
    \caption{Demonstration of the median-filtering step to isolate emission lines and continuum. \textbf{Top panel}: The kernel for median filtering with a full width of 31 pixels and a central ``hole'' of 5 pixels. \textbf{Bottom panel}: The 1D extracted spectra(gray), continuum (green), and emission line (blue) data from a simulated QSO spectrum. The emission line locations are marked with red dashed lines.}
    \label{fig: detrend}
\end{figure}

We begin the redshift measurement procedure by detrending the continuum using an empirical median filtering technique with a boxcar-like kernel, following the approach in \cite{2023ApJ...950...66K}. The primary kernel for this median filtering has a length of 31 pixels with a central ``hole'' of 5 pixels (about 3000 km $\text{s}^{-1}$) to avoid the self-subtraction around broad emission lines. Figure \ref{fig: detrend} presents the kernel shape and continuum subtracting results of one spectrum. While quasars exhibit a wide range of emission line widths, the continuum may sometimes be overestimated in the wings of emission lines. However, the resulting absorption-like features in these wings do not interfere with the subsequent emission line detection.

After removing the continuum, we detect the emission line peaks in the following steps. First, the detrended spectrum is divided into three segments corresponding to the GU, GV, and GI bands, with a 300\AA~overlap between adjacent segments to ensure that peaks near segment boundaries are not missed. Within each segment, we identify the data points where the flux exceeds both 1.5 times the standard deviation of the segment data and twice the flux uncertainty. These identified points are then grouped, and the center of each emission line is determined. Such a process may introduce fake emission line peaks, but these false detections could not match the actual emission lines we build up in Table \ref{tab: el_infor}.

Finally, We apply the cross-correlation techniques to match the detected peaks to the QSO emission line maps from \cite{2001AJ....122..549V}. To avoid the misidentifications for weaker lines, we add more prominent weights for stronger emission lines. We also consider a higher tolerance for the emission lines with possible blue-shift or variance profiles caused by nearby broad absorption lines. To quantify the quality of the redshift measurement, we consider the following definitions. 
\begin{itemize}
    \item [-] Redshift measurements with two or more peaks identified in the spectra. In such cases, the emission line could be identified by the wavelength ratio between multi-peaks. With the emission line identifications, we could obtain the redshift by comparing the wavelength of the observation peak centers and the rest-frame emission lines. The final redshift is calculated as the weighted average of the redshifts of all detected emission lines, with weights being the reciprocals of their tolerances. These sources are labeled as class 2;
    \item [-] Only one emission line detected in the spectra. These sources may have some emission lines too weak to be detected or false peaks at the edge of the grism, and are labeled as class 1;
    \item [-] No emission line detected in the spectra. This situation is mainly caused by noisy spectra or falsely classified sources that do not have emission line features. These sources are labeled as class 0.
\end{itemize}

\begin{figure}[htbp]
    \centering
    \includegraphics[width=0.47\textwidth]{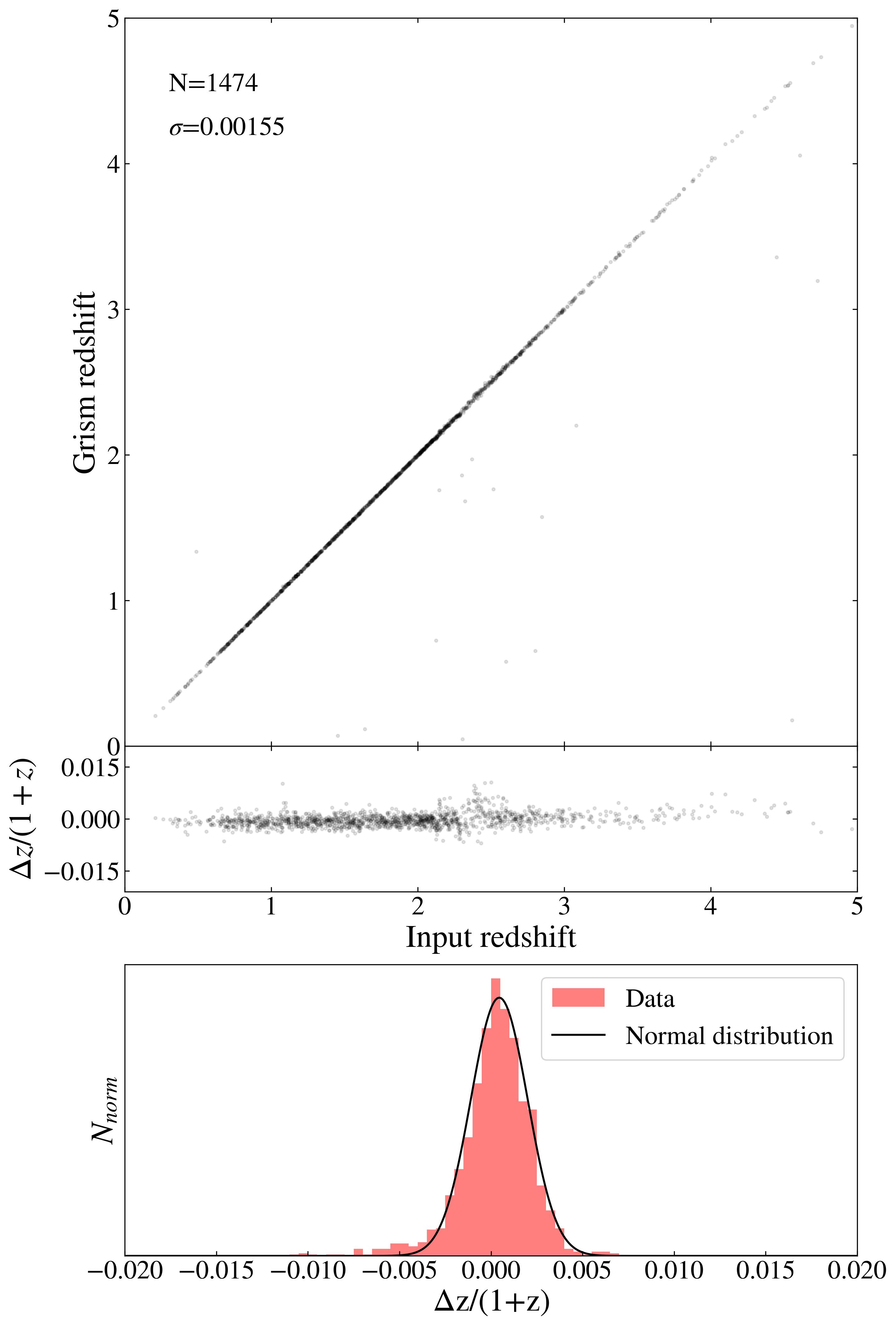}
    \caption{Redshift comparison between the input catalog and the emission line matching results. \textbf{Top panel}: Redshifts comparison for objects with two or more emission line detections. The number of objects and the $\sigma_{\mathrm{NMAD}}$ value are noted in the top left corner. \textbf{Middle panel}: $\Delta z/(1+z)$ as a function of redshift. \textbf{Bottom panel}: The distribution of $\Delta z/(1+z)$.}
    \label{fig: z_compare}
\end{figure}

Using the redshift identification procedure, 1474/1644 (89.7\%), 150/1644 (9.6\%), and 12/1644 (0.7\%) sources are classfied as the class 2, 1, 0, respectively. Figure \ref{fig: z_compare} shows the comparison results between the redshifts given by the grism spectra pipeline for class 2 sources and those in the input catalog. We adopt the normalized median absolute deviation of errors ($\sigma_{\mathrm{NMAD}}$), and the outlier fraction ($f_{\mathrm{c}}$) as the evaluation metrics for the redshift estimation. These metrics are defined as follows:

\begin{gather}
    \sigma_{\mathrm{NMAD}} = 1.48 \times \mathrm{median} \left(\left| \frac{\Delta z- \mathrm{median}(\Delta z)}{1+z} \right|\right) \\
    f_{\mathrm{c}} = \frac{1}{n} \times \mathrm{count}\left(\left| \frac{\Delta z}{1+z} \right| > 0.01\right),
\end{gather}

The class 2 sources reach a $\sigma_{\mathrm{NMAD}}$ value of 0.0016, which is consistent with our estimation in Section \ref{sec: discuss_redshift}. The $f_{\mathrm{c}}$ is 1.5\%(22/1474) for class 2 sources which comes from the emission line misidentification.


\begin{deluxetable}{lccc}
\tablecaption{QSO emission lines used for redshift determination \label{tab: el_infor}}
\tabletypesize{\footnotesize}
\tablehead{Name & Wavelength & Tolerance & Weight}
\startdata
{Ly $\alpha$} & 1215.24 & 0.1 & 9.0 \\
{Si~\textsc{IV}} & 1396.76 & 0.05 & 1.0 \\
{C~\textsc{IV}} & 1549.48 & 0.12 & 8.0 \\
{C~\textsc{III}]} & 1908.73 & 0.06 & 7.0 \\
{Mg~\textsc{II}} & 2799.11 & 0.05 & 8.0 \\
{O~\textsc{II}} & 3727.09 & 0.05 & 6.0 \\
{H $\gamma$} & 4341.68 & 0.05 & 3.0 \\
{H $\beta$} & 4862.68 & 0.05 & 4.0 \\
{[O~\textsc{III}]} & 4960.29 & 0.03 & 2.0 \\
{[O~\textsc{III}]} & 5008.24 & 0.03 & 2.0 \\
{H $\alpha$} & 6564.61 & 0.05 & 8.0 \\
\enddata
\end{deluxetable}

\section{Fitting Results of Mock Spectra} \label{sec: fitting}

\begin{deluxetable}{lccc}
\tablecaption{The emission line fitting parameters of QSOFITMORE \label{tab: el_fitting}}
\tablehead{Line complex & Fitting range($\rm{\AA}$) & Line & $n_{Gauss}$}
\startdata
Lyman~$\alpha$ & 1150-1290 & Lyman~$\alpha$~broad & 1 \\
 &  & Lyman~$\alpha$~narrow & 1 \\
 &  & N~\textsc{V} & 1 \\ 
C~\textsc{IV} & 1500–1600 & C~\textsc{IV} broad & 3 \\
 &  & He~\textsc{II}1640 & 1 \\
 &  & [O~\textsc{III}] 1663 & 1 \\
Si~\textsc{IV} & 1290-1450 & Si~\textsc{IV}1396 & 1 \\
 &  & Si~\textsc{IV}1402 & 1 \\
C~\textsc{III}] & 1850–1970 & C~\textsc{III}] broad & 2 \\
Mg~\textsc{II} & 2700–2900 & Mg~\textsc{II} broad & 2 \\
 &  & Mg~\textsc{II} narrow & 1 \\
H$\gamma$ & 4250-4440 & H$\gamma$ & 1 \\ 
\hb & 4640-5100 & \hb~broad & 3 \\
 &  & \hb~narrow & 1 \\
 &  & [O~\textsc{III}]4959 & 1 \\
 &  & [O~\textsc{III}]5007 & 1 \\
 &  & He~\textsc{II}4687 & 1 \\
\ha & 6400–6800 &\ \ha~broad & 3 \\
 &  & \ha~narrow & 1 \\
 &  & [N~\textsc{II}]6549 & 1 \\
 &  & [N~\textsc{II}]6585 & 1 \\
 &  & [S~\textsc{II}]6718 & 1 \\
 &  & [S~\textsc{II}]6732 & 1 
\enddata
\end{deluxetable}

In the previous sections, we constructed 1452 pure QSO samples with good redshift determinations. In this section, we present the fitting and the physical parameter estimation results of these sources.

\subsection{Black Hole Mass and Eddington Ratio Estimation of QSOs} \label{sec: 5.1}
\begin{figure*}[htbp]
    \centering
    \includegraphics[width=0.90\textwidth]{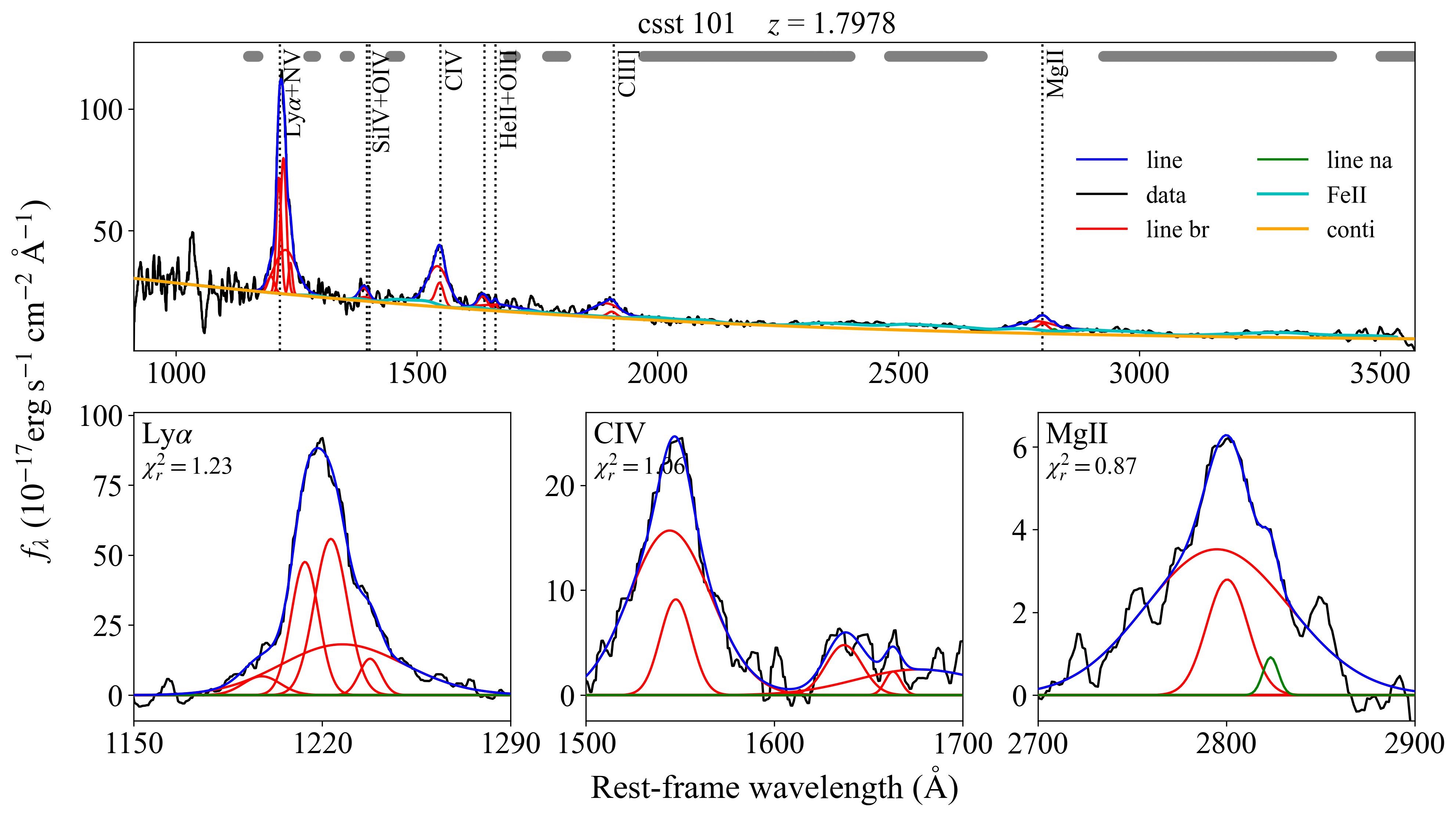}
    \caption{QSOFITMORE fitting results of a QSO's 1D spectrum in our simulation. The black lines denote the total extracted spectrum, the yellow lines denote the continuum, the cyan lines denote the Fe~\textsc{II} templates ($f_{Fe~\textsc{II}}$), the blue lines denote the total flux of different emission lines, and the red lines denote each single Gaussian components.}
    \label{fig: fitting-example}
\end{figure*}

To characterize the statistical properties of the simulated 1D spectra, we employ QSOFITMORE \citep[version 1.1.0;][]{yuming_fu_2021_5810042}, a wrapper package built upon PyQSOFit \citep[][]{2018ascl.soft09008G}, to fit the spectra with accurately determined redshifts as described in the previous section. We refer to \citet{2011ApJS..194...45S, 2019ApJS..241...34S} and \citet{2022ApJS..261...32F} for the details of the fitting procedure. 
All the spectra are corrected for Galactic extinction with the dust map \citep{2014A&A...571A..11P} and the extinction law from \cite{2019ApJ...877..116W}. For sources with redshifts below 1.16, the spectra undergo a decomposition into a host-galaxy component and a QSO component using the principal component analysis (PCA) method \citep{2004AJ....128..585Y, 2004AJ....128.2603Y} implemented in PyQSOFit. Subsequently, we employ a more flexible third-order polynomial function and a Fe~\textsc{II} model ($f_{Fe\textsc{II}}$) to fit a pseudo-continuum ($f_{cont}$) of the QSO spectrum after masking the emission lines. The emission-line components are then fitted with multiple Gaussian profiles following the subtraction of the continuum component from the spectrum. Table \ref{tab: el_fitting} outlines the parameters used for line fitting. Furthermore, we utilize the Monte Carlo (MC) method to estimate the uncertainties associated with the measured quantities. Figure \ref{fig: fitting-example} shows a typical fitting example of an extracted QSO spectrum from the simulation.

To ensure sufficient data points for multi-Gaussian fitting for emission lines, we interpolate the spectra using the \verb|specutils.resample| function to get a data point density equivalent to a resolution of 1200. Moreover, since the narrow line widths become more extended due to the spectral space blending effect in the slitless spectroscopy, we extend the upper limit of the fitting for narrow line widths to the same limit of the corresponding broad lines while maintaining the lower limit as the standard setting of QSOFITMORE.

\begin{figure*}[htbp]
    \centering
    \includegraphics[width=1\textwidth]{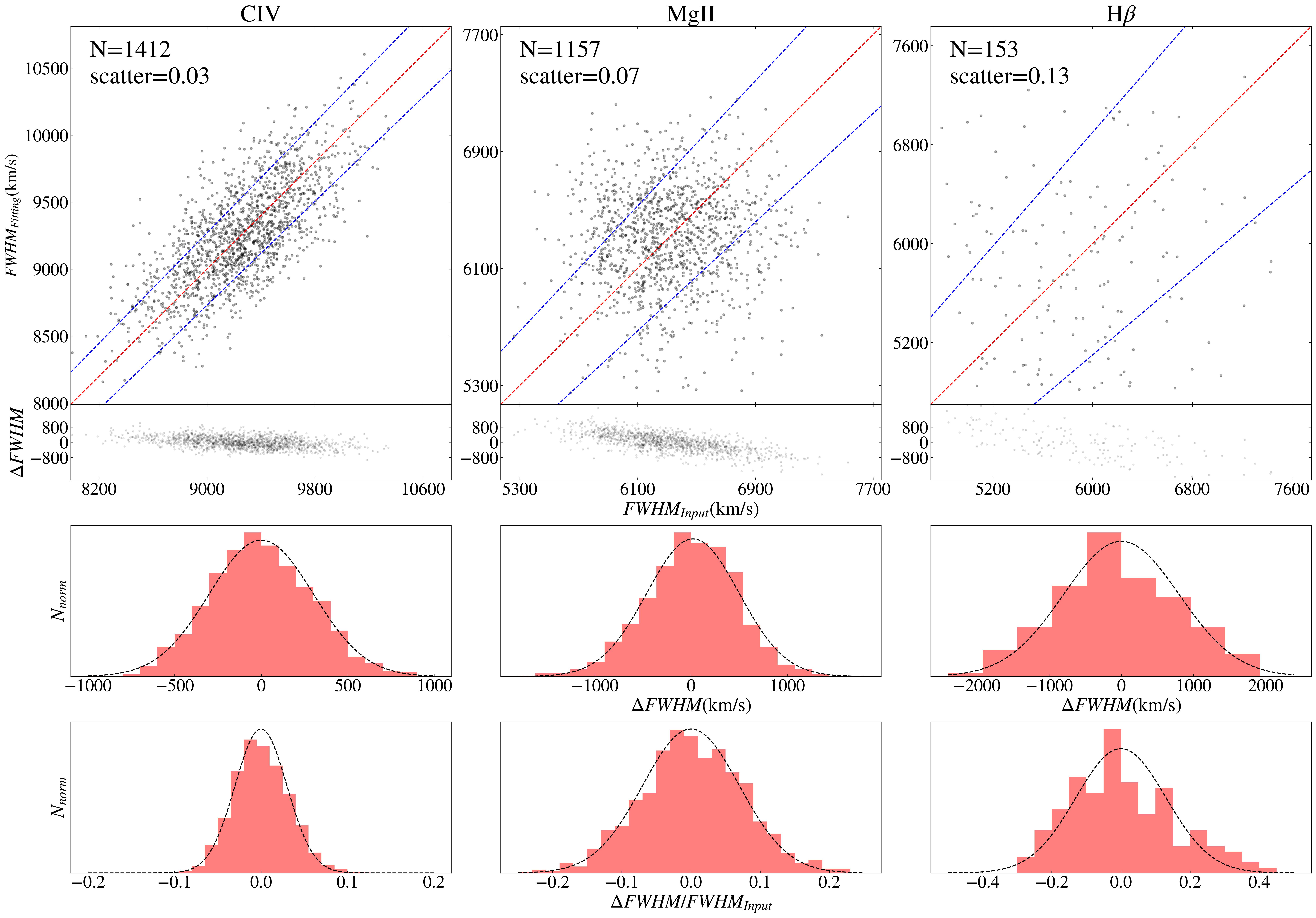}
    \caption{Comparison between the emission line widths from the 1D extracted spectra and the input catalog. Panels from left to right show the results of the C~\textsc{IV}, Mg~\textsc{II}, and \hb~emission lines. \textbf{Top panels}: FWHM comparison between different lines. The total number of samples and the standard error are noted in the top left corner in each panel. \textbf{Middle panels}: FWHM difference ($\Delta$FWHM) as a function of $\text{FWHM}_{\text{input}}$ and the $\Delta$FWHM distributions. \textbf{Bottom panels}: Distributions of $\Delta$FWHM for each emission line. The black dash line represents the Gaussian distribution which $\sigma$ equals to the standard error in the top panel.}
    \label{fig: FWHM_compare}
\end{figure*}

Upon the completion of the QSOFITMORE fitting, the output full width at half maximum (FWHM) parameter tends to be significantly larger than the input value due to the convolution of the line profile with the PSF pattern. Therefore, we fit the FWHM values using the linear combination of the Gaussian width ($\sigma$), FWHM, and monochromatic luminosity ($L_{1350}$, $L_{3000}$, $L_{5100}$ corresponding to C~\textsc{IV}, Mg~\textsc{II}, \hb) outputted by QSOFITMORE for half of the samples. Then we apply such linear relation to the entire sample set to get the calibrated FWHM value. Figure \ref{fig: FWHM_compare} compares the calibrated FWHM and the input spectra for the C~\textsc{IV}, Mg~\textsc{II}, and \hb~emission lines.

With the continuum luminosity serving as a proxy for the broad-line region size \citep[i.e., the R-L relation; e.g.,][]{2000ApJ...533..631K, 2004A&A...424..793W, 2006ApJ...644..133B, 2016ApJ...825..126D} and the broad-line width as a proxy for the virial velocity, we estimate the single-epoch virial black hole masses (\mbh) of our simulation sample using the empirical mass-scaling relations calibrated with the reverberation mapping (RM) masses \citep[e.g.,][]{2004MNRAS.352.1390M, 2006ApJ...641..689V, 2009ApJ...707.1334W}. To derive the virial black hole masses, we adopt the H$\beta$-based estimator from \citet{2006ApJ...641..689V}, the Mg~\textsc{II}-based estimator from \citet{2009ApJ...707.1334W}, and the C~\textsc{IV}-based estimator from \citet{2006ApJ...641..689V} as described below:

\begin{equation}
\begin{split}
    \mathrm{l}&\mathrm{og}(M_{\mathrm{BH}}/M_{\odot})  \\
    &=\mathrm{log} \left[\left(\frac{\mathrm{FWHM}(\mathrm{H}\beta)}{\mathrm{km~s^{-1}}}\right)^2\left(\frac{L_{5100}}{10^{44}\mathrm{erg~s^{-1}}}\right)^{0.5}\right]+0.91
\end{split}
\end{equation}
\begin{equation}
\begin{split}
    \mathrm{l}&\mathrm{og}(M_{\mathrm{BH}}/M_{\odot})  \\
    &=\mathrm{log} \left[\left(\frac{\mathrm{FWHM}(\mathrm{Mg~\textsc{II}})}{\mathrm{km~s^{-1}}}\right)^{1.51}\left(\frac{L_{3000}}{10^{44}\mathrm{erg~s^{-1}}}\right)^{0.5}\right]+2.60
\end{split}
\end{equation}
\begin{equation}
\begin{split}
    \mathrm{l}&\mathrm{og}(M_{\mathrm{BH}}/M_{\odot})  \\
    &=\mathrm{log} \left[\left(\frac{\mathrm{FWHM}(\mathrm{C~\textsc{IV}})}{\mathrm{km~s^{-1}}}\right)^{2}\left(\frac{L_{1350}}{10^{44}\mathrm{erg~s^{-1}}}\right)^{0.53}\right]+0.66
\end{split}
\end{equation}

The uncertainties in the measurements of black hole masses are calculated from the propagation of MC errors of the line widths and the monochromatic continuum luminosities. Additionally, we estimate the bolometric luminosity ($L_{bol}$) and its error computed from $L_{5100}$, $L_{3000}$, $L_{1350}$ using the spectral fits and bolometric corrections ($\mathrm{BC}_{5100}$ = 9.26, $\mathrm{BC}_{3000}$ = 5.15, and $\mathrm{BC}_{1350}$ = 3.81) derived from the composite spectral energy distribution (SED) in \citet{2006ApJS..166..470R}. We then utilize \mbh~and $L_{bol}$ to compute the Eddington ratio ($\lambda_{\rm{Edd}}$) of these QSOs by:

\begin{equation}
    \lambda_{\rm{Edd}}=\frac{L_{bol}}{L_{\rm{Edd}}}=\frac{L_{bol}}{1.26\times10^{38}\rm{erg/s}}\frac{M_{\odot}}{M_{\rm{BH}}}
\end{equation}

\begin{figure}[htbp]
    \centering
    \includegraphics[width=0.47\textwidth]{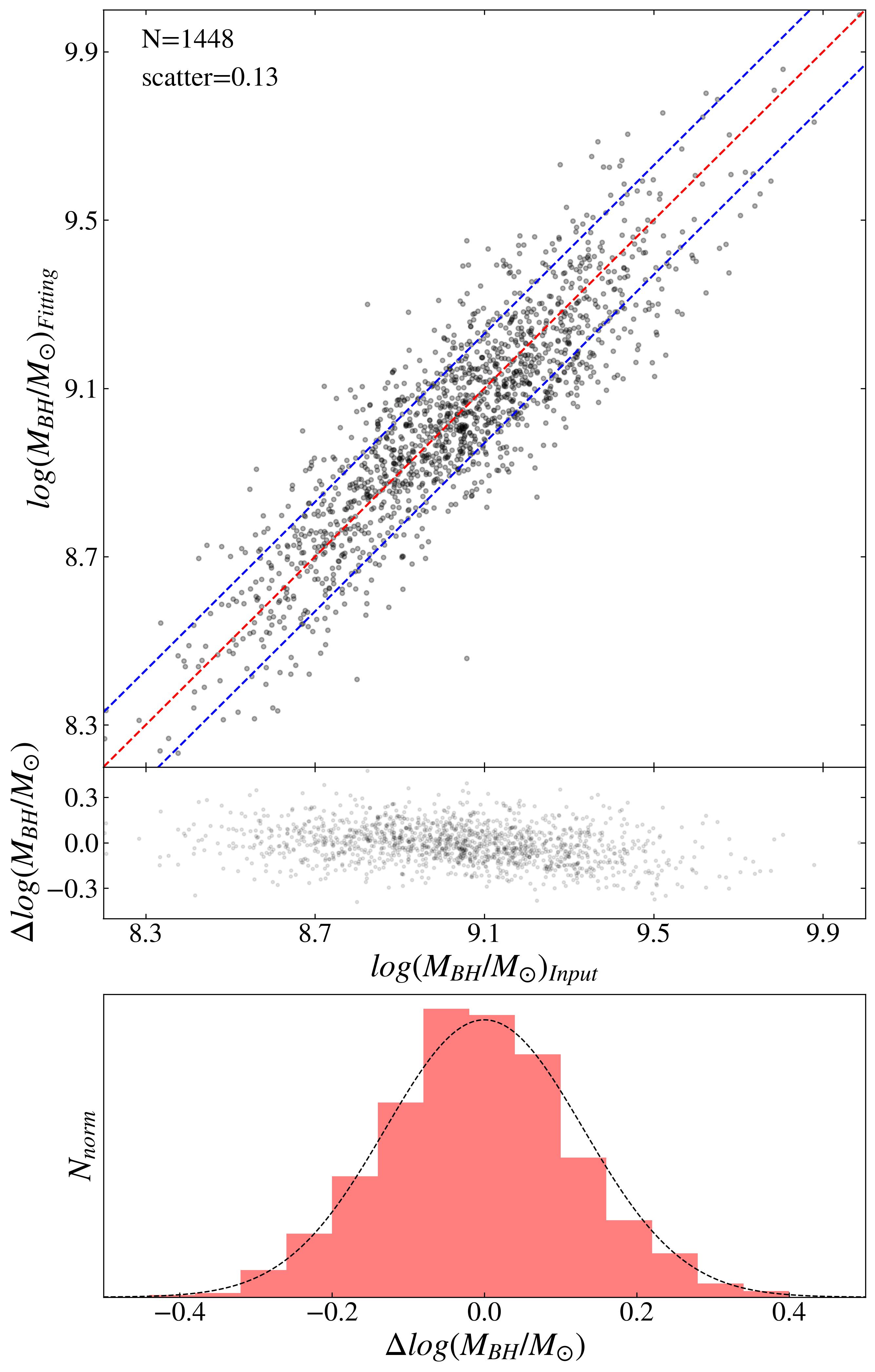}
    \caption{Comparison between the $M_{BH}$ estimation results from the C~\textsc{IV}, Mg~\textsc{II}, and \hb~emission line and the input catalog. Different panels have the similar meaning as in Figure \ref{fig: FWHM_compare}}
    \label{fig: MBH_compare}
\end{figure}

\begin{figure}[htbp]
    \centering
    \includegraphics[width=0.47\textwidth]{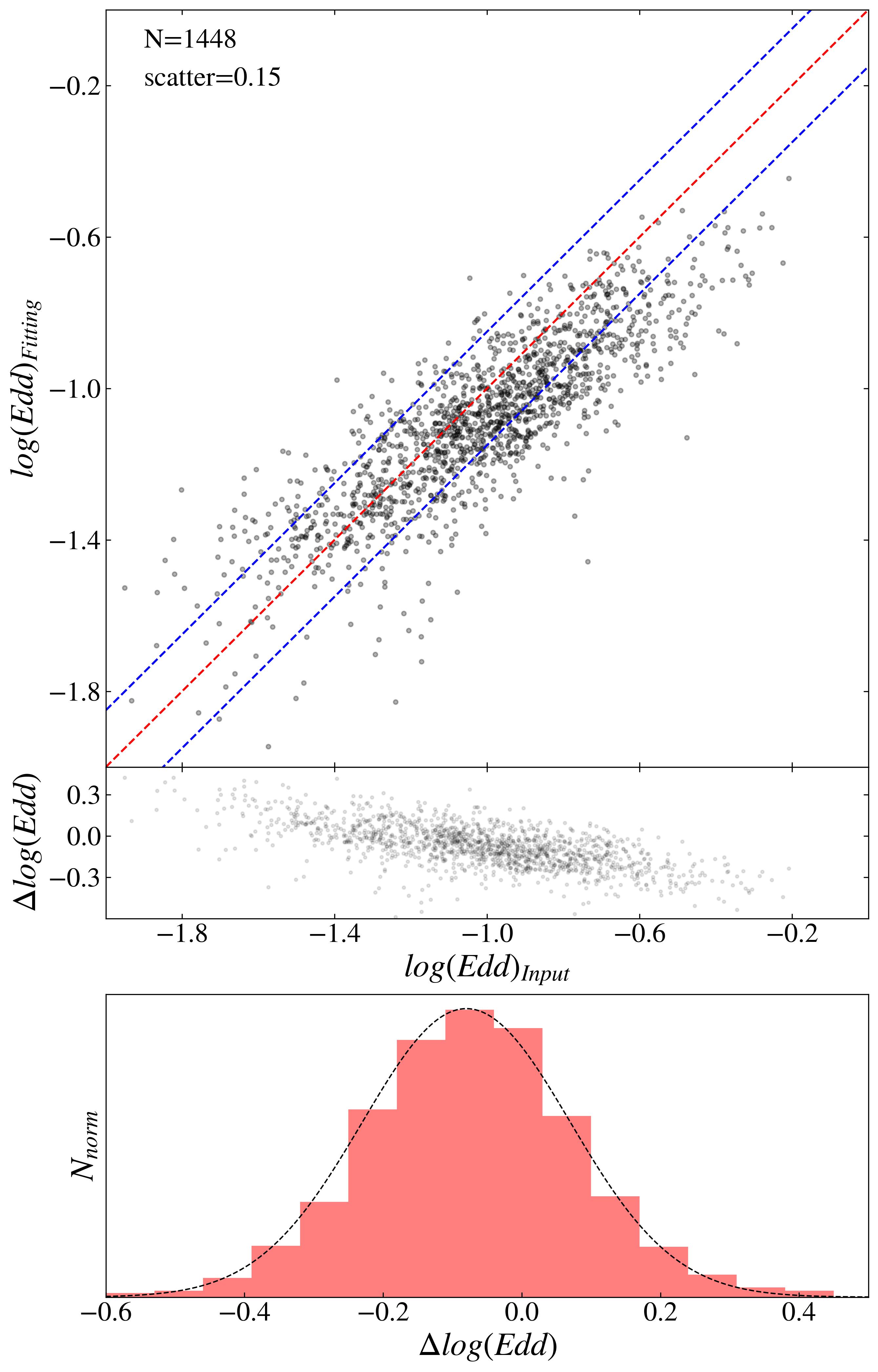}
    \caption{Comparison between the $\lambda_{\rm{Edd}}$ estimation results from 1D extracted spectra and the input catalog. Different panels have the similar meaning as in Figure \ref{fig: FWHM_compare}}
    \label{fig: Edd_compare}
\end{figure}

Figure \ref{fig: MBH_compare} and Figure \ref{fig: Edd_compare} illustrate the comparison between the input and computed results for \mbh~and $\lambda_{\rm{Edd}}$. Overall, the standard error of the calibrated FWHM is approximately 10\%. Through the fitting results of \hb, Mg~\textsc{II}, and C~\textsc{IV} emission lines, \mbh~and $\lambda_{\rm{Edd}}$ are obtained for 1448 samples, and the dispersions of \mbh~and $\lambda_{\rm{Edd}}$ estimations are 0.13 and 0.15 dex, respectively. It is worth noting that the dispersion of \mbh~and $\lambda_{\rm{Edd}}$ estimations does not account for the systematic uncertainty of the single-epoch virial method.

\subsection{Metallicity Estimation of QSOs} \label{sec: 5.2}
QSOs can provide valuable insights into the evolution of metallicity across cosmic time \citep[e.g.,][]{2007AJ....134.1150J, 2011ApJ...739...56D, 2021ApJ...907L...1W}. Previous studies have indicated that the metallicity of the broad-line region (BLR) in QSOs has not significantly evolved since redshift 7 \citep[e.g.,][]{2007ApJ...669...32K, 2020ApJ...898..105O, 2020ApJ...905...51S}. Besides, the correlations may exist between BLR metallicities and specific physical properties of QSOs. For example, \citet{2006A&A...447..157N} found a significant positive correlation between BLR metallicity and QSO luminosity, while \citet{2012ApJ...751L..23W} identified a correlation between BLR metallicity and outflows. Investigating such relationships in a larger QSO sample from the CSST survey could enhance our understanding of QSO evolution. Therefore, we plan to investigate the accuracy of metallicity diagnosis from the emission line ratio using our simulation sample.

\begin{figure*}[htbp]
    \centering
    \includegraphics[width=0.7\textwidth]{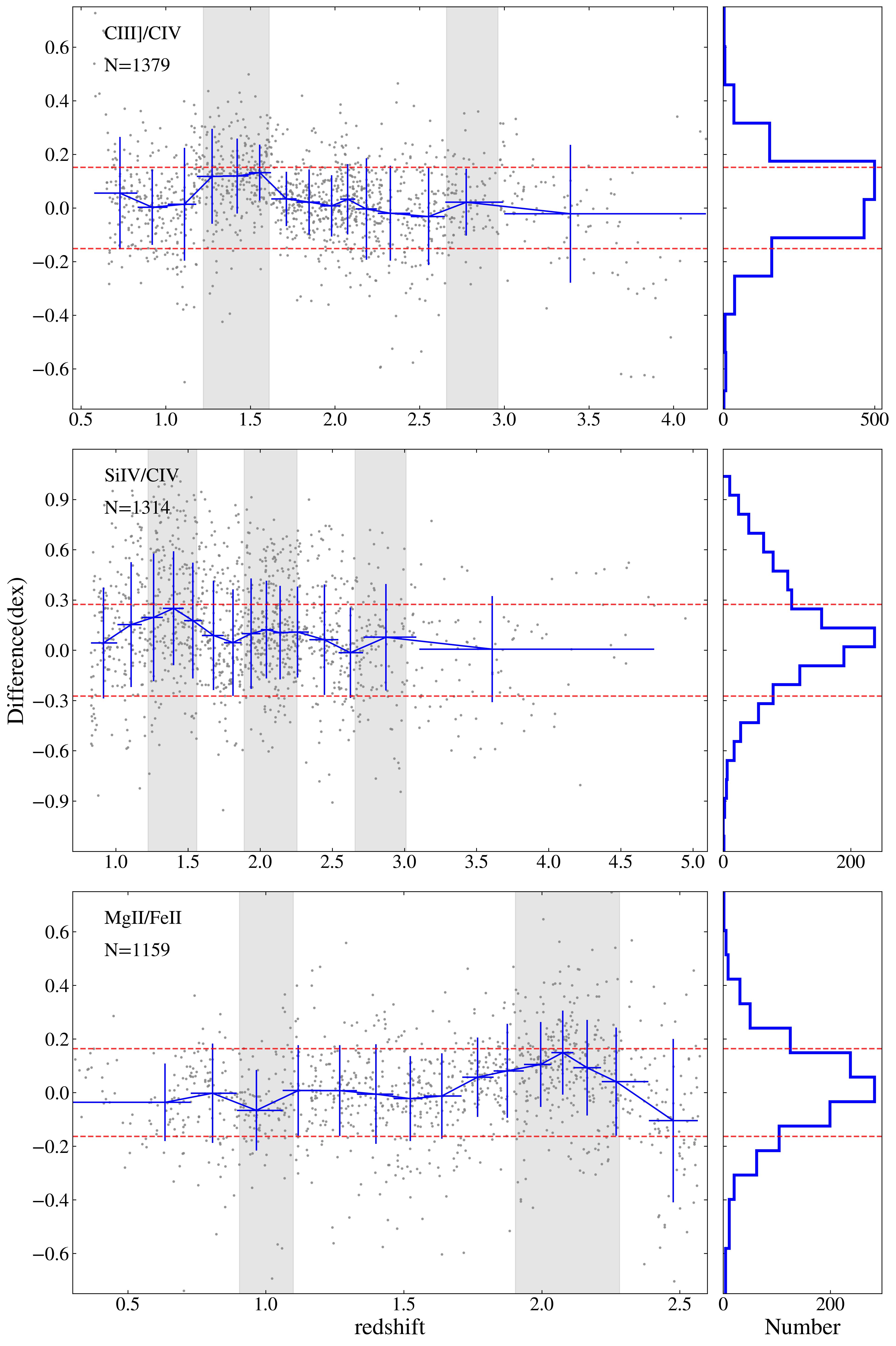}
    \caption{Comparison between the line ratio estimation results from the 1D extracted spectra and the line ratio in the input catalog at different redshift. The panels from top to bottom display different metallicity diagnosis results, and the name is noted on the upper-left corner with the total number of objects. \textbf{Left panels}: The blue curve shows the median value (represented by the center of the crosses) and the scatter (represented by the vertical bar) for different redshift bins (represented by the horizontal bar), and background grey dots show the results for individual sources. The red dashed lines represent the typical error of line ratio estimation. \textbf{Right panels}: Distribution of the line ratio estimation difference.}
    \label{fig: metallicity}
\end{figure*}

In this work, all spectral quantities are measured from the QSOFITMORE best-fit models. The Fe~\textsc{II} flux is computed by integrating the best-fit Fe~\textsc{II} model over the rest-frame wavelength range 2200–3090 Å \citep[the same as][]{2011ApJS..194...45S, 2011ApJ...739...56D}. We sum the multiple Gaussians integrated over the full line profile for broad emission-line fluxes. The difference between the fitting results of the input and extracted 1D spectra are shown in Fig \ref{fig: metallicity}, Table \ref{tab: metallicity} summarizes the results of the bias and scatter of different metallicity diagnoses.

\begin{deluxetable}{lcc}
\tablecaption{The results of matellicity diagnostics \label{tab: metallicity}} 
\tabletypesize{\footnotesize}
\tablehead{\colhead{Line complex} & \colhead{Bias} & \colhead{Scatter}}
\startdata
C~\textsc{III}]/C~\textsc{IV} & 0.033 & 0.178\\
Si~\textsc{IV}/C~\textsc{IV} & 0.335 & 0.093\\
Fe~\textsc{II}/Mg~\textsc{II} & 0.015 & 0.191\\
\enddata
\end{deluxetable}

\subsubsection{UV Line Flux Ratios and Metallicity}
The measurement of QSO BLR metallicities primarily relies on the rest-frame UV broad emission lines. Photoionization models suggest that the UV line flux ratios, such as (Si~\textsc{IV}+O~\textsc{IV}])/C~\textsc{IV}, (C~\textsc{III}]+Si~\textsc{IV}, Al~\textsc{III}/C~\textsc{IV}, N~\textsc{V}/C~\textsc{IV}, He~\textsc{II}/C~\textsc{IV}, and N~\textsc{V}/He~\textsc{II}, can be used to infer metallicities \citep[e.g.,][]{2002ApJ...564..592H, 2006A&A...447..157N}. Therefore, examining how these line ratios deviate in the slitless spectra compared to the standard long-slit spectra is essential. Our analysis focuses on the main metallicity diagnostic lines, including C~\textsc{IV}, He~\textsc{II}, Si~\textsc{IV}, N~\textsc{V}, C~\textsc{III}], and Mg~\textsc{II}. We exclude weaker lines, such as Al~\textsc{III}, as they are too faint to be identified in the slitless spectra.


Figure \ref{fig: metallicity} illustrates the deviation of various line ratios as a function of redshift, compiled from the slitless spectra. Here are the comparison results:
\begin{itemize}
    \item [-] C~\textsc{III}]/C~\textsc{IV}: C~\textsc{III}] and C~\textsc{IV} are both strong, broad emission lines of QSOs; thus, the SNR of the emission line flux is relatively higher. As shown in the top panels of Figure \ref{fig: metallicity}, the C~\textsc{III}]/C~\textsc{IV} ratio has a scatter with about 0.17 dex compared to the input spectra. The C~\textsc{III}]/C~\textsc{IV} ratio has deviations at specific redshift where C~\textsc{III}] or C~\textsc{IV} emission lines in the GU and GV overlapped low sensitivity area (at z=1.2-1.6), and C~\textsc{IV} emission lines in the GV and GI overlapped low sensitivity area (at z=2.65-2.95).
    \item [-]Si~\textsc{IV}/C~\textsc{IV}: Si~\textsc{IV} is a weaker line compare to C~\textsc{III}], thus the SNR is smaller. As shown in the middle panels of in Figure \ref{fig: metallicity}, the Si~\textsc{IV}/C~\textsc{IV} ratio has a larger scatter of up to 0.3 dex compared to the input spectra. Si~\textsc{IV}/C~\textsc{IV} ratio has deviations in larger redshift range, including C~\textsc{IV} emission lines in the GU and GV overlapped low sensitivity area (at z=1.3-1.6), C~\textsc{IV} emission lines in the GV and GI overlapped low sensitivity area (at z=2.65-2.95), and Si~\textsc{IV} in the GU and GV overlapped low sensitivity area (at z=1.9-2.25).
    \item [-]Other weak emission lines: We also test other weak emission lines like Al~\textsc{III} or N~\textsc{V}, but these weak lines are hard to be identified campared with the strong, broad emission lines because of the low spectral resolution. Thus, the total flux of these emission lines is unreliable for diagnosing metallicity. 
\end{itemize}

Despite the significant scatter, the median values of these line flux ratios are generally consistent with those obtained from the input spectra across most of the redshift range. The median values and the typical standard scatter of our sample are summarized in Table \ref{tab: metallicity}. Different line ratios are affected by the junction positions of the GU, GV, and GI grisms at various redshifts, and the scatter in weak line ratios is too large for reliable metallicity estimation.

\subsubsection{Fe~\textsc{II}/Mg~\textsc{II}}
Core-collapse supernovae and Type Ia supernovae can produce different ratios of Fe and Mg elements \citep[][]{1995MNRAS.277..945T, 1997NuPhA.616...79N,1997NuPhA.621..467N}. Therefore, the Fe~\textsc{II}/Mg~\textsc{II} ratio, used as a first-order approximation of the Fe/Mg abundance ratio, can provide essential constraints on the star formation history of QSO host galaxies. Moreover, a high Fe~\textsc{II}/Mg~\textsc{II} ratio may have a positive correlation with the Eddington ratio, indicating the need for further black hole mass corrections for super-Eddington black holes \citep[][Pan et al. in prep]{2016ApJ...825..126D}.

Figure \ref{fig: metallicity} compares the Fe~\textsc{II}/Mg~\textsc{II} ratios (see the bottom panel) from the slitless spectra with those from the input spectra. There is no significant bias in the Fe~\textsc{II}/Mg~\textsc{II} distribution between the slitless and standard spectra across most of the redshift range, with a 0.19 dex scatter compared to the input spectra. Moreover, when the 2000 Å peak of the Fe II template aligns with the junctions of the GU/GV and GV/GI grisms, the Fe II flux has a more significant deviation due to the artifacts produced by the low-sensitivity areas. That causes a more considerable bias in the redshift range of 0.9-1.1 and 1.9-2.3. Our sample's median values and typical standard scatters are also summarized in Table \ref{tab: metallicity}.

\section{Discussion} \label{sec: discuss}

\subsection{Redshift Accuracy through Emission Line Matching and Machine Learning method} \label{sec: discuss_redshift}
\begin{figure}[htbp]
    \centering
    \includegraphics[width=0.49\textwidth]{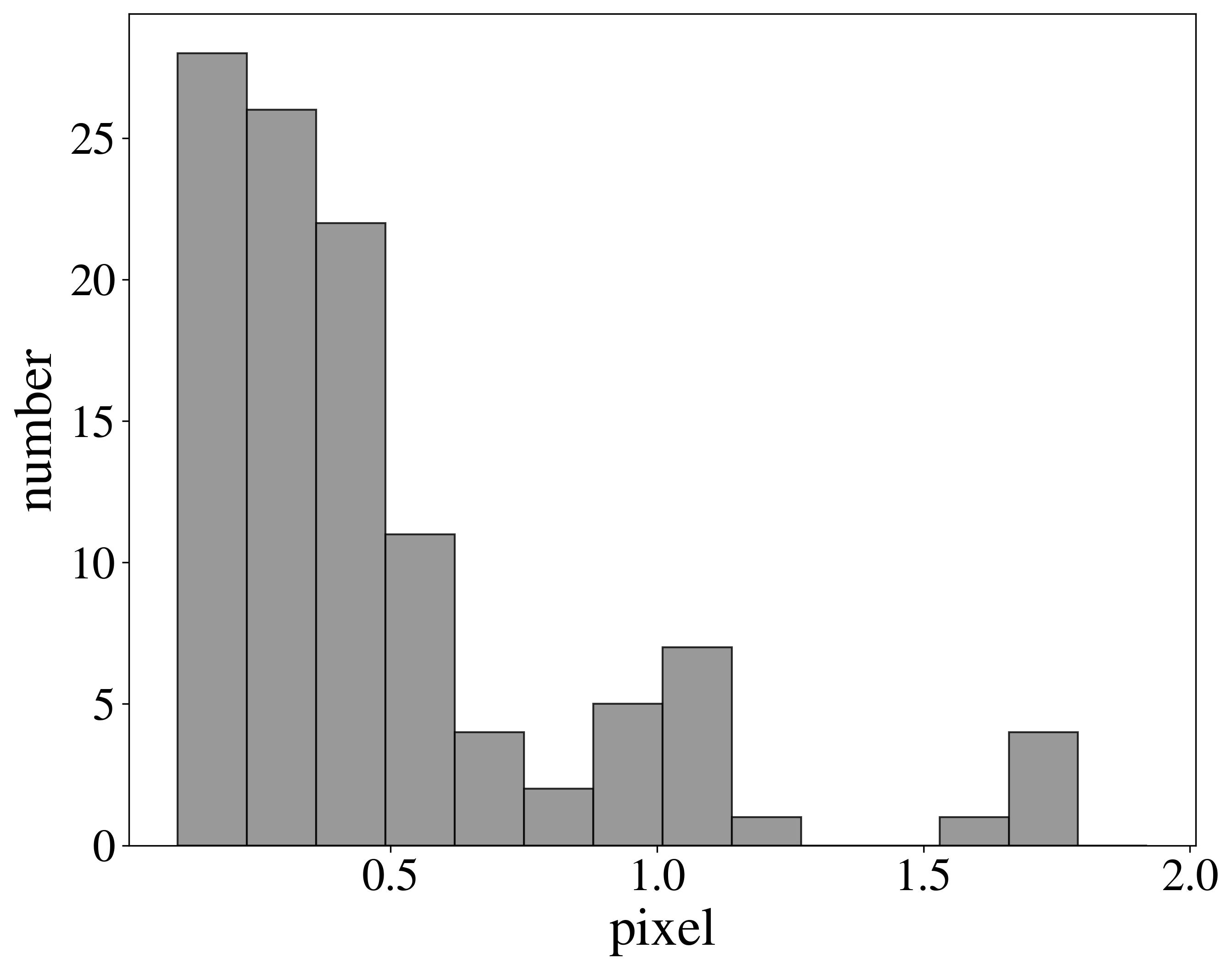}
    \caption{The distance distribution between the actual positions and the extracted positions of sources in the GU, GV, and GI images. The gap of most sources is within 1 pixel. A few sources with relatively large distances are faint sources in the GU grism.}
    \label{fig: loc_cali}
\end{figure}

In this paper, we determine the redshifts of QSOs by matching emission lines. Since we use a combination of multiple line centers to identify the final redshift, the difference between the final and the input redshift mainly comes from the uncertainty in the wavelength calibration. We assume an ideal polynomial dispersion function in our simulation, therefore, the uncertainty in the wavelength calibration comes from the offset in the source extraction position. Figure \ref{fig: loc_cali} compares the location differences between the input and output in one of the simulation orbits. Most sources with enough SNR have the location differences of about 0.3-0.4 pixels. The differences in location calibration introduce a redshift scatter of approximately 0.002, which aligns with the redshift determination results in Section \ref{sec: redshift}. In actual cases, the calibration file is fitted using the observation data and may lead to additional systematic effects. Thus, we estimate that the actual redshift scatter is about 0.003, similar to the HST calibration result \citep{2016ApJS..225...27M}.


Other limitations exist in determining the redshifts of sources through the emission line matching. For example, in the simulation, about 10\% of the sources have fewer than two emission lines detected, making it very difficult to identify the emission lines and, therefore, determine the redshift. We also consider using a machine learning model with a similar procedure in Section \ref{sec: distinguish} for redshift training (with a 3:1 training-to-testing ratio). In general, NN can provide redshifts for all sources, with an $\sigma_{\mathrm{NMAD}}$ scatter about 0.032. Such accuracy is not enough for the QSOFITMORE program since the redshift difference will cause a large blue or red shift in the emission line. However, with the accumulation of CSST observational data, the deep learning procedure should be able to achieve better results than the emission line matching procedure \citep[QuasarNet,][]{2018arXiv180809955B}.

\subsection{SNR of the Extracted Spectra}
\begin{figure*}[htbp]
    \centering
    \includegraphics[width=0.95\textwidth]{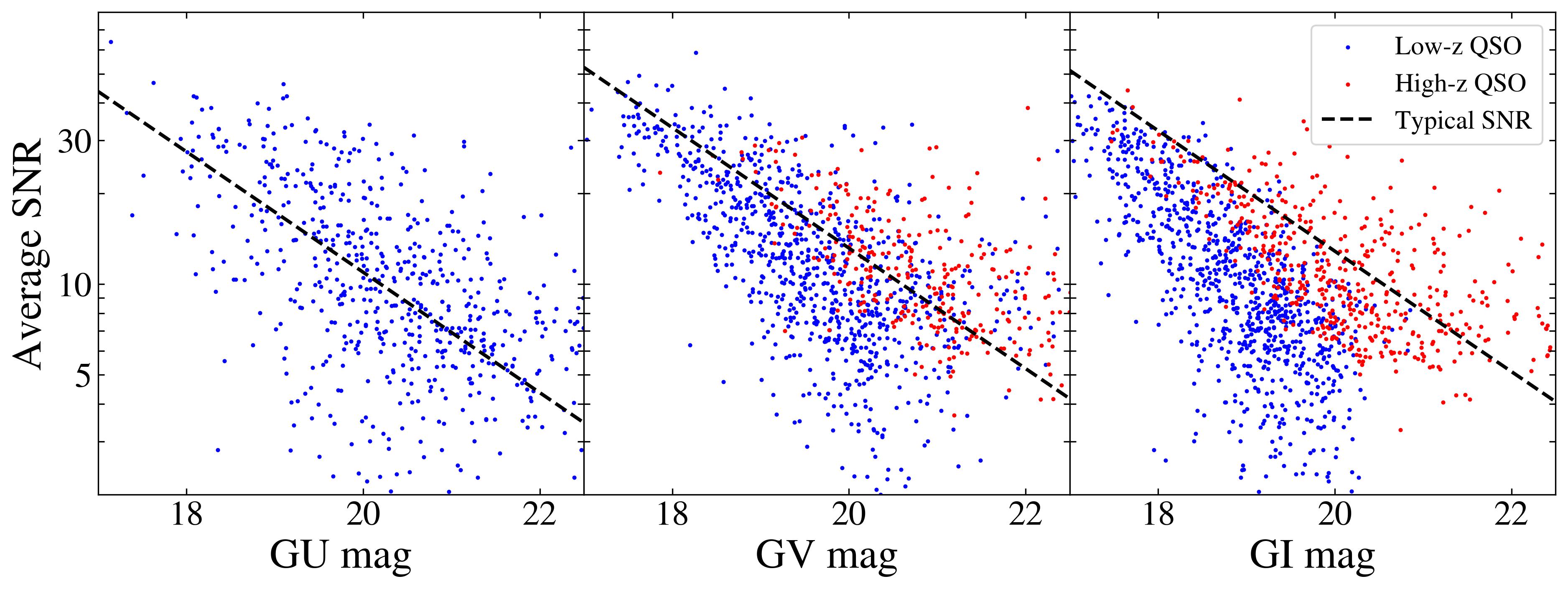}
    \caption{The average SNR calculated from the extracted QSO spectra across different grisms. Panels from left to right shows the results of GU, GV, and GI grism. The blue and red dots represent low-z and high-z QSOs, separated by a redshift of 2, respectively. The dashed black line represented the designed parameter of the CSST telescope.}
    \label{fig: disscuss_flux}
\end{figure*}

Figure \ref{fig: disscuss_flux} illustrates the average SNR across different grisms calculated from the extracted QSO spectra. The SNR is consistent with each grism's ideal design of limited magnitude (dashed black line in Figure \ref{fig: disscuss_flux}). The scatter is primarily caused by the continuum and emission line shape differences due to spatial contamination effects. Several sources exhibit a higher SNR due to strong Lyman~$\alpha$~emission lines. Low-z QSOs with host galaxy components in the GV and GI grisms require a larger extraction aperture. Thus, the 2D extraction introduces more background noise and shows a more significant SNR decrease than high-z QSOs.

\subsection{Impact of Missing Observation and Variability on Classification and Redshift Measurement}
Although CSST's grisms have a wide wavelength coverage, they can do observations separately in the GU, GV, and GI bands. Considering the unstable situation in space, it is necessary to discuss the impacts of missing one or two grisms on QSO classification and redshift measurement. 

We use the same simulated precedure but remove one or two grism's observation data, then classify the extracted spectra using the same neural network structure described in Section \ref{sec: distinguish}. As shown in Table \ref{tab: accuracy_class}, missing one of the grism observation will reduce the success rate of the classification since the wavelength coverage is not large enough. The remaining GV and GI grisms will be better since it has more extensive wavelength coverage and could identify the Lyman~$\alpha$~break feature at a redshift larger than 2. Only one grism band remaining will make the classification more difficult since less information remains. The GU grism group seems to have the highest success rate, but it is based on the bias of the remaining sample. High-z QSOs will have no signal in the GU grism and will not be detected, while low-z QSOs will have strong Lyman~$\alpha$~and C~\textsc{IV} lines in the GU grism, marking a promising feature to be detected. It is worth noticing that due to the vast difference in populations between QSOs and galaxies, only a few percent of the mis-classification of galaxies could lead to a disastrous drop in the accuracy of the QSO samples. 

The absence of the grism observation also results in the loss of the emission line features, making redshift determination more challenging. We use the QSOs classified in Section \ref{sec: redshift} as the basic sample and measure their redshifts by emission line matching under different grism absence situation. The results are presented in Table \ref{tab: redshift_discuss}. Although $f_{\mathrm{c}}$ is still low in all groups, which means the redshift results are still conceivable when matching with two or more emission lines in the extracted spectra, the total detection numbers of two or more emission lines drops significantly since some strong emission lines may be absent in the grism detection.

Due to CSST's unique observational strategy, there is an observational interval between capturing the GU, GV, and GI spectra of a source. Consequently, QSOs may exhibit variability between the observations in different grism observations. To account for this situation, we simulate the spectra with a typical variability of 0.1 mag \citep{1996A&A...306..395C, 2012ApJ...758..104Z} and classify the QSOs using the same method as in Section \ref{sec: distinguish}. As summarized in Table \ref{tab: accuracy_class}, the variability only slightly influence the classification results.

\begin{deluxetable}{lccccc}
\tablecaption{The accuracy of classification in different groups \label{tab: accuracy_class}} 
\tabletypesize{\footnotesize}
\tablehead{\colhead{Data group} & \colhead{Accuracy} & \colhead{Precision} & \colhead{Recall}  & \colhead{$F_{1}$} & \colhead{MCC}}
\startdata
All data & 0.9921 & 0.9920 & 0.9920 & 0.9920 & 0.9890 \\
Missing GU & 0.9427 & 0.9399 & 0.9326 & 0.9343 & 0.9079 \\
Missing GV & 0.9771 & 0.9773 & 0.9771 & 0.9771 & 0.9687 \\
Missing GI & 0.9461 & 0.9488 & 0.9488 & 0.9487 & 0.9295 \\
Only GU & 0.940 & 0.9473 & 0.9443 & 0.9446 & 0.9226 \\
Only GV & 0.8861 & 0.9044 & 0.9065 & 0.9044 & 0.8674 \\
Only GI & 0.8145 & 0.8445 & 0.8325 & 0.8355 & 0.7729 \\
Variability & 0.9917 & 0.991 & 0.9908 & 0.9909 & 0.9875
\enddata
\end{deluxetable}

\begin{deluxetable}{lcccc}
\tablecaption{The emission line matching results and $f_{\mathrm{c}}$ in redshift determination for different cases of missing grism observation \label{tab: redshift_discuss}} 
\tabletypesize{\footnotesize}
\tablehead{\colhead{Data group} & \colhead{No EL} & \colhead{One EL} & \colhead{Two or more EL}  & \colhead{ $f_{\mathrm{c}}$}}
\startdata
All data & 0.73\% & 9.58\% & 89.39\% & 0.27\% \\
Missing GU & 8.54\% & 37.65\% & 53.33\% & 0.34\% \\
Missing GV & 16.23\% & 27.33\% & 51.04\% & 0.56\% \\
Missing GI & 1.88\% & 16.13\% & 81.69\% & 0.45\% \\
Only GU & 25.24\% & 42.88\% & 31.87\% & 0.57\% \\
Only GV & 16.55\% & 50.79\% & 32.30\% & 1.31\% \\
Only GI & 52.60\% & 38.17\% & 6.51\% & 0.91\% \\
\enddata
\end{deluxetable}

\subsection{Impact of Contamination in Slitless Spectroscopy}
In our simulations, the source density was relatively sparse, so we didn't account for the effects of contamination in slitless spectroscopy. Here, we discuss the impact of nearby source contamination and self-contamination on source classification and redshift measurement.

As estimated in Section \ref{sec: distinguish}, approximately 10\% of quasars are affected by contamination from nearby sources. Additionally, faint AGNs can experience self-contamination caused by their host galaxies. In such cases, the continuum features in the 1D spectra can be significantly distorted, leading to challenges in classification. Forward modeling to reconstruct the intrinsic spectra of target sources is a commonly used method with HST, including object-based techniques \citep[e.g., Grizli,][]{2019ascl.soft05001B} and pixel-based techniques \citep[e.g., Linear,][]{2018PASP..130c4501R}. However, these methods are time-consuming and may be challenging to apply to the large datasets from the CSST wide-field survey. Another potential approach involves training convolutional neural networks (CNNs) directly on the 2D slitless spectroscopic data. Such an approach would require a larger sample of known sources and might also necessitate simulated data to build up different contamination scenarios for training.

Beyond the classification, contamination can also affect redshift determination. Additional emission lines may appear in the extracted spectrum when nearby sources exhibit strong emission lines (e.g. emission line galaxies), or the zeroth-order spectra of the contaminated source overlap with the target spectra. Such additional peaks could confuse the redshift determination of QSOs. To mitigate these effects, data near zeroth-order spectra should be masked during the extraction. Additionally, photometric redshift estimation from the photometry from NUV to y bands can help identify the belonging sources of each emission line and associate them with the correct spectral lines.

As the survey progresses and more data are accumulated, deep learning techniques could also be explored. By combining the 2D imaging and slitless spectroscopic data, such methods may expect to automatically identify spectral line positions and associate them with the corresponding sources, providing a more robust solution to the contamination in slitless spectroscopy.

\subsection{Future Prospective}
In addition to provide the slitless spectroscopy data, CSST will also engage a multi-band image survey which is two magnitudes deeper than the grism observations with the PSF FWHM at 0.15 arcsec. These photometry data could also provide precious information to select a deeper QSO sample. In our future work, we will discuss the selection precision, the effectiveness of different selection methods, and the accuracy of the AGN host decomposition in details.

Slitless spectra could be regarded as low-resolution IFU data for the emission line. The multi-band image can also help build a continuum model for the 2D slitless spectra. By constraining the continuum model, we could get the 2D emission line map for the host galaxies of AGN to calculate the metallicity \citep[e.g.,][]{2023ApJ...943...37B}. However, this technique is more difficult in the CSST database because of the different observation times, depths, and resolutions between the image and grism observations. Thus, we still need further tests to get a synthetic reference images for grisms by combining the image information and the PSF of the image and grism.

\section{Summary} \label{sec: summary}
The CSST wide field survey of 17500 $\text{deg}^2$ observations could help us find a large, complete QSO sample. This work presents the standard pipeline for the GU, GV, and GI grisms operating in the spectral range 2550-10000$\rm{\AA}$. We build up the simulated data of QSOs, galaxies, and stars using the CSST Cycle 6 simulation procedure. Starting from the spectra extraction, we classify the QSO sample, determine their redshift, and discuss the slitless spectra fitting results. We summarise the main results as follows:

\begin{itemize}
    \item [-] Using the convolutional neural network (CNN), we effectively classified QSOs from galaxies and stars based on slitless spectra, achieving a success rate of up to 99\% (see Section \ref{sec: distinguish}). About 0.9 million new QSOs in the redshift range of 0-5 can be identified through grism observations alone under such high classification accuracy, providing an ideal database to evaluate the completeness of QSO surveys using other QSO selection methods.
    \item [-] Redshift measurements are robust for most QSOs because the broad wavelength coverage can detect multiple strong emission lines. Using the emission line detection and matching procedure, 90\% of the classified QSO samples have two or more reliable line detections, providing 98.5\% with $\Delta z/(1+z)<0.01$ precision (see Section \ref{sec: redshift} and Figure \ref{fig: z_compare}).
    \item [-] We fit the parameters of emission lines using the QSOFITMORE procedure. Since the emission line profiles suffer from the low resolution and spatial convolutional effects, we calibrate the FWHM value using the fitting results of $\sigma$, FWHM, and monochromatic luminosity parameters. After the calibration, we compare \mbh~and $\lambda_{\rm{Edd}}$ deriving from the slitless spectra and the input catalog. Our results show a 0.13 dex scatter in \mbh~estimation and 0.15 dex scatter in $\lambda_{\rm{Edd}}$ estimation with almost no bias (see Section \ref{sec: 5.1}, Figure \ref{fig: MBH_compare} and Figure \ref{fig: Edd_compare}).
    \item [-] We discuss the uncertainty of BLR metallicity diagnoses of QSOs. Despite some specific redshift ranges are biased due to the low sensitivity in the overlapped regions of GU, GV, and GI grisms, C~\textsc{III}]/C~\textsc{IV} and Fe~\textsc{II}/Mg~\textsc{II} ratios have scatters of about 0.18 dex. In comparison, Si~\textsc{IV}/C~\textsc{IV} has a larger scatter with 0.3 dex, and other emission lines like Al~\textsc{III} or N~\textsc{V} are too weak to get reliable fitting results (see Section \ref{sec: 5.2}).
\end{itemize}

In the future, we plan to substantially extend our preliminary studies on the CSST QSO survey. By applying our method to a larger sample of QSOs, we aim to analyze the constraints on the QSO luminosity function within the CSST's wide survey area. Additionally, we will investigate the technical aspects, including the impact of contamination from neighboring spectra in high-density regions and the identifications of hot or defective pixels in the extracted spectra. The techniques we developed for the CSST slitless QSO survey may be also helpful to other space missions with grism facility such as Euclid \citep{2017SPIE10563E..4WR,2024arXiv240518126E} and Roman \citep{2019BAAS...51c.413R}.

\acknowledgments
We thank Zhiwei Pan, Haicheng Feng and Xin Zhang for helpful discussions. We acknowledge the support of the National Key R\&D Program of China (2022YFF0503401). We thank the support from the National Science Foundation of China (12133001 and 11927804) and the science research grant from the China Manned Space Project with No. CMS-CSST-2021-A06.

\bibliography{CSST_simu_2024}{}
\bibliographystyle{aasjournal}
\end{document}